\documentclass[prc,twocolumn,nofootinbib,superscriptaddress]{revtex4}
\usepackage{amsmath,amssymb,epsfig,bm,mathrsfs}
\usepackage{feynmp,slashed,color,mathrsfs,nicefrac,exscale}
\usepackage[colorlinks=true,linkcolor=blue,citecolor=blue,urlcolor=blue]{hyperref}
\newcommand{\bea}{\begin{eqnarray}}
\newcommand{\eea}{\end{eqnarray}}
\newcommand{\be}{\begin{equation}}
\newcommand{\ee}{\end{equation}}

\newcommand{\vecp}{{\bm p}}

\newcommand{\Tr}{{\rm Tr}}

\newcommand{\ie}{{\it i.e.}}
\newcommand{\eg}{{\it e.g.}}

\def\XXint#1#2#3{{\setbox0=\hbox{$#1{#2#3}{\int}$}
     \vcenter{\hbox{$#2#3$}}\kern-.5\wd0}}

\def\apj{ApJ}%
\def\pra{Phys. Rev. A}%
\def\prc{Phys. Rev. C}%
\def\prd{Phys. Rev. D}%
\def\physrep{Phys. Rep.}%
\definecolor{red}{rgb}{0.8,0,0}
\definecolor{violet}{rgb}{0.4,0,0.4}
\definecolor{green}{rgb}{0,0.5,0.0}
\definecolor{navy}{rgb}{0.0,0.0,0.6}
\definecolor{orange}{rgb}{0.8,0.2,0.0}
   
\begin{document}
\title{
Bulk viscosity of  two-flavor quark matter from the Kubo formalism
}
\author{Arus Harutyunyan}\thanks{arus@th.physik.uni-frankfurt.de}
\affiliation{Institute for Theoretical 
   Physics, Goethe-University, D-60438 Frankfurt-Main, Germany 
}

\author{Armen Sedrakian} \thanks{sedrakian@fias.uni-frankfurt.de
} 
\affiliation{Frankfurt Institute for Advanced Studies, D-60438
  Frankfurt-Main, Germany }

\begin{abstract}
  We study the bulk viscosity of quark matter in the strong coupling
  regime within the two-flavor Nambu--Jona-Lasinio model.  The
  dispersive effects that lead to nonzero bulk viscosity arise from
  quark-meson fluctuations above the Mott transition temperature,
  where meson decay into two quarks is kinematically allowed.  We
  adopt the Kubo-Zubarev formalism and compute the equilibrium
  imaginary-time correlation function for pressure in the $O(1/N_c)$
  power counting scheme. The bulk viscosity of matter is expressed in
  terms of the Lorentz components of the quark spectral function and
  includes multiloop contributions which arise via resummation of
  infinite geometrical series of loop diagrams. We show that the
  multiloop contributions dominate the single-loop contribution close
  to the Mott line, whereas at high temperatures the one-loop
  contribution is dominant.  The multiloop bulk viscosity dominates
  the shear viscosity close to the Mott temperature by factors 5 to
  20, but with increasing temperature the shear viscosity becomes the
  dominant dissipation mechanism of stresses as the one-loop
  contribution becomes the main source of bulk viscosity.
\end{abstract}

\maketitle

\section{Introduction}
\label{sec:introduction}

The transport coefficients of quark-gluon plasma continue to attract
significant attention as they are key inputs in the hydrodynamical
description of heavy-ion collisions at the energies of the
Relativistic Heavy Ion Collider (RHIC) and Large Hadron Collider
(LHC). The data on elliptic flow in the heavy-ion collisions can be
well described by a low value of the shear viscosity $\eta$ of the
fluid, with the ratio of the shear viscosity to the entropy density $s$ being
close to the lower bound placed by the uncertainty principle~\cite{1985PhRvD..31...53D} and
conjectured from AdS/CFT duality arguments~\cite{2005PhRvL..94k1601K}.

The role of the bulk viscosity, which describes the dissipation in the
case where pressure falls out of equilibrium on uniform expansion or
contraction of a statistical ensemble, is more subtle.  As it is well known,
bulk viscosity vanishes in a number of cases, \eg, for ultrarelativistic
and nonrelativistic gases interacting weakly with local forces via
binary collisions~\cite{LIFSHITZ19811,1971ApJ...168..175W}.

The bulk viscosity $\zeta$ of quark-gluon plasma is small in the
perturbative regime~\cite{1985NuPhB.250..666H,2006PhRvD..74h5021A,2008JHEP...09..015M,2013PhRvD..87c6002C},
but was found to be large close to the critical temperature of the
chiral phase transition. For example, lattice simulations of the pure
gluodynamcis close the critical temperature predict
$\zeta/s\sim 1$~\cite{2008PhRvL.100p2001M},
where $s$ is the entropy density, and it is expected that
$\zeta$ becomes singular at the critical point of second order phase
transition~\cite{2006PhRvC..74a4901P}.  Values of $\zeta/s\sim 1$
affect the description of data in heavy-ion
collisions~\cite{2012PhRvC..85d4909D} and can lead to a breakdown of
the fluid description via onset of
cavitation~\cite{2010JHEP...03..018R}.

Controlled computations of the bulk viscosity exist in perturbative
QCD on the basis of kinetic theory of relativistic
quarks~\cite{2006PhRvD..74h5021A,2008JHEP...09..015M,2013PhRvD..87c6002C}. In
the strongly coupled regime various approximate methods were applied,
including QCD sum rules in combination with the lattice data on the
QCD equation
state~\cite{2008JHEP...09..015M,2008PhLB..663..217K,Aarts:2007va}
and quasiparticle Boltzmann
transport~\cite{2010NuPhA.832...62S,2011PhRvC..83a4906C,2011PhRvD..84i4025C,2012EPJC...72.1873D,Marty:2013ita}. Some
strongly coupled systems can exhibit zero bulk viscosity if the scale or, more generally, the conformal symmetry is intact. This is the
case, for example, in atomic Fermi gases in the unitary
limit~\cite{2006PhRvA..74e3604W,2007PhRvL..98b0604S,2011AnPhy.326..770E,2013PhRvL.111l0603D},
but not in the QCD and QCD-inspired theories when the conformal
symmetry is broken by the quark mass terms and/or by dimensionful
regularization of the ultraviolet divergences. This is indeed the case
in the Nambu--Jona-Lasinio (NJL)  model of low-energy QCD that we will
utilize below.

A nonperturbative method to compute the transport coefficients of
quark-gluon plasma close to the chiral phase transition is based on
the Kubo-Zubarev
formalism~\cite{1957JPSJ...12..570K,zubarev1997statistical}, with the
correlators computed from the quark spectral function derived from the
NJL model in conjunction with the $1/N_c$ diagrammatic
expansion~\cite{1994PhRvC..49.3283Q}. This approach has been applied
extensively to compute the shear viscosity of quark
plasma~\cite{2008JPhG...35c5003I,
  2008EPJA...38...97A,LW14,LKW15,LangDiss,2016PhRvC..93d5205G}, but
there exist only a few computations of the bulk
viscosity~\cite{2014ChPhC..38e4101X,2016PhRvC..93d5205G} in this
regime.

In this work we extend the previous study of the transport
coefficients of two-flavor quark mater within the Kubo-Zubarev
formalism and NJL model~\cite{2017arXiv170204291H} to compute the bulk
viscosity of quark plasma close to the critical line of the chiral
phase transition. We specifically argue that the one-loop result for
the correlation function of quarks, which arises in the leading order
of $1/N_c$ expansion, cannot be applied in the case of bulk viscosity
and a resummation of infinite series is required. As a consequence,
our results are substantially different from those obtained previously from the
one-loop computations.

For completeness we point out that the bulk viscosity of dense and cold
QCD was extensively discussed in the context of compact stars and
strange stars because it is the dominant dissipation mechanism to damp
the unstable Rossby waves ($r$-modes)
~\cite{2007PhRvC..75e5209A,2007PhRvD..75f5016S,2007PhRvD..75l5004S,2007PhRvD..75g4016D,2008JPhG...35k5007A,
  2010PhRvD..81d5015H,2011NJPh...13d5018S,2016PhRvD..94l3010B}. In
this regime of QCD the bulk viscosity is dominated by the weak
interaction process like $\beta$-decays of quarks
$d\to u + e + \bar\nu$ or nonleptonic weak process in three-flavor
quark matter $u+d\to u+s$.  The time scales associated
  with the weak processes are much larger than the collisional
  time scale. The situation is an analogue of the case of bulk
  viscosity of fluids undergoing chemical reactions on time scales much
  larger than the collisional time scale, which may lead to large bulk
  viscosity, as shown long ago by
  Mandelstam and Leontovich~\cite{ML}. This contribution to the bulk
  viscosity is called ``soft-mode'' contribution, because it is described
  by the response of the system to small frequency
  perturbations~\cite{2015PhRvC..91b5805K}. As we are interested here
  in the hydrodynamical description of heavy-ion collisions, which
  have characteristic time-scales much shorter than the weak
  time scale, we will not discuss weak processes.  Slow ``chemical
  equilibration'' processes may play a role in the bulk viscosity in
  the multicomponent environment in heavy-ion collisions, but are
  beyond the scope of this work.

The paper is organized as follows. Section~\ref{sec:Kubo_bulk} starts
from the Kubo-Zubarev formula for the bulk viscosity and expresses it
in terms of the Lorentz components of the quark spectral function. In
Sec.~\ref{sec:QSpectralFunctions} we summarize the results of
Ref.~\cite{2017arXiv170204291H} for the quark spectral function, in
the case where the dispersive effects arise from the quark-meson
fluctuations. Our numerical results for the bulk viscosity are
collected in Sec.~\ref{sec:results}.  Section~\ref{sec:conclusions}
provides a short summary of our results.  Appendix \ref{app:A}
describes the details of the computation of the bulk viscosity beyond
one-loop approximation. In Appendix \ref{app:B} we discuss the
thermodynamics of the model and derive a number of relations that are
required for the computation of the bulk viscosity. We use the natural
(Gaussian) units with $\hbar= c = k_B  = 1$, and the metric signature $(1,-1,-1,-1)$.

\section{Kubo formula for bulk viscosity}
\label{sec:Kubo_bulk}

We consider two-flavor quark matter described by the NJL-Lagrangian of 
the form   
\be\label{eq:lagrangian}
\mathcal{L}=\bar\psi(i\slashed \partial-m_0)\psi+
\frac{G}{2}\left[(\bar\psi\psi)^2+
(\bar\psi i\gamma_5\bm\tau\psi)^2\right],
\ee 
where $\psi=(u,d)^T$ is the iso-doublet quark field, $m_0=5.5$ MeV is
the current-quark mass, $G=10.1$ GeV$^{-2}$ is the effective
four-fermion coupling constant and $\bm\tau$ is the vector of Pauli
isospin matrices. This Lagrangian describes four-fermion contact
scalar-isoscalar and pseudoscalar-isovector interactions between
quarks with the corresponding bare vertices $\Gamma^0_{s}=1$ and
$\Gamma^0_{ps}=i\bm\tau\gamma_5$.  The symmetrized energy momentum
tensor is given in the standard fashion by
\be\label{eq:energymom}
T_{\mu\nu}=\frac{i}{2}(\bar\psi\gamma_{\mu}
\partial_{\nu}\psi +\bar\psi\gamma_{\nu}
\partial_{\mu}\psi)-g_{\mu\nu}\mathcal{L}.
\ee 
The net particle current is given by
\bea\label{eq:current}
N_\mu=\bar{\psi}\gamma_\mu\psi,
\eea
which is the only conserved current in the
case of isospin-symmetric quark matter, \ie, quark matter described by a single chemical potential for both flavors.

The Kubo and Zubarev formalisms relate the transport properties of
material to different types of {\it equilibrium} correlation functions
of an ensemble~\cite{zubarev1997statistical,1957JPSJ...12..570K},
which in turn can be computed from equilibrium many-body techniques.

The bulk (second) viscosity within the Kubo-Zubarev formalism 
is given by~\cite{1987NuPhB.280..716H,2011AnPhy.326.3075H}
\be\label{eq:bulk}
\zeta =-\frac{d}{d\omega}\rm {Im}\Pi^R_\zeta(\omega)\bigg|_{\omega=0},
\ee
where the relevant two-point correlation function is given by 
\bea\label{eq:corp1}
\Pi^{R}_\zeta(\omega) 
= -i\int_{0}^{\infty}dt\ e^{i\omega t}\int d\bm r\langle
\left[\hat{p}^*(\bm r,t),\hat{p}^*(0)\right]\rangle_{0},
\eea
with
\bea\label{eq:p_star}
\hat{p}^*(\bm r,t)&=&\hat{p}(\bm r,t)-\gamma \hat{\epsilon}(\bm r,t)
-\delta \hat{n}(\bm r,t)\nonumber\\
&=&\frac{1}{3}T_{ii}(\bm r,t) 
-\gamma T_{00}(\bm r,t)-\delta N_0(\bm r,t).
\eea
Here $\hat{p}$, $\hat{\epsilon}$ and $\hat{n}$ are operators of the pressure, the energy density and
the particle number density, respectively; the second line uses the
relation between these quantities and energy-momentum tensor and particle number current in the fluid rest frame; $\gamma$ and $\delta$ are thermodynamic quantities and are given by
\bea\label{eq:gammadelta}
\gamma =\bigg(\frac{\partial p}
{\partial\epsilon}\bigg)_n,\quad 
\delta =\bigg(\frac{\partial p}
{\partial n}\bigg)_\epsilon. 
\eea 
The last term in Eq.~\eqref{eq:p_star} is 
present only at finite chemical potentials;
see Appendix~\ref{app:B} for details.

\begin{figure}[t] 
\begin{center}
\includegraphics[width=6cm,keepaspectratio]{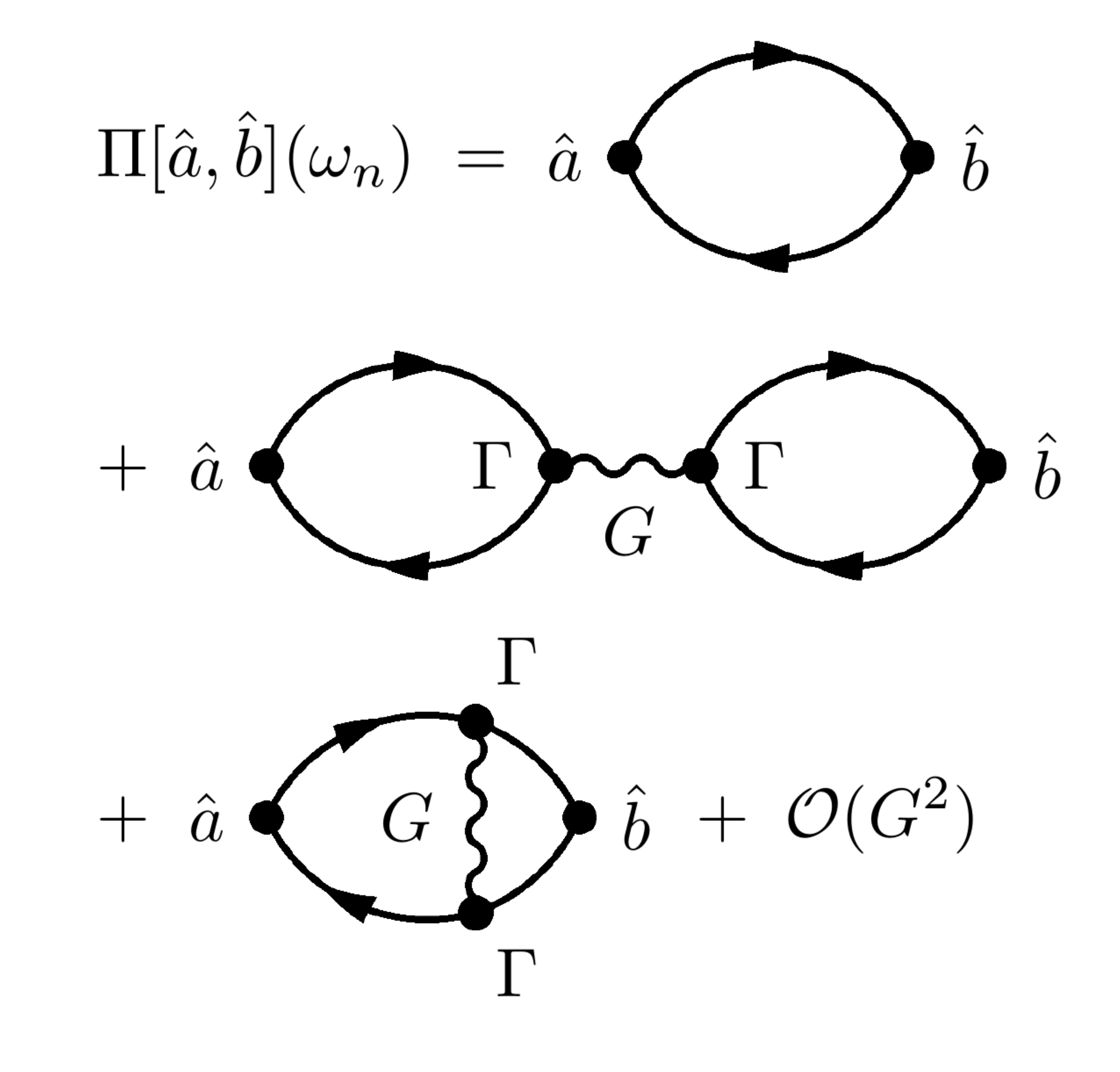}
\caption{ Contributions to the two-point correlation functions from 
  ${\cal O}(N_c^1)$ (first and second lines) and   ${\cal O}(N_c^0)$ (the 
third line) diagrams which are either of zeroth or first order in the coupling constant $G$. }
\label{fig:loops_ab} 
\end{center}
\end{figure}

Inserting Eq.~\eqref{eq:p_star} into Eq.~\eqref{eq:corp1} we obtain a set of two-point correlation functions of the generic form
\bea\label{eq:rings_ab}
\Pi[\hat{a},\hat{b}](\omega_n) =
\int_{0}^{\beta}\!\! d\tau e^{i\omega_n\tau}
\int d\bm r
\langle {\cal T}_\tau(\bar\psi\hat{a}\psi\Big\vert_{(\bm r,\tau)},
\bar\psi\hat{b}\psi\Big\vert_0)\rangle_0,\nonumber\\
\eea 
where we switched to the imaginary-time Matsubara formalism by means
of the substitutions $t\to -i\tau$, $\partial_t\to i\partial_\tau$.
In Eq.~\eqref{eq:rings_ab} $\omega_n=2\pi nT$, $n=0,\pm 1,\ldots$ is a
bosonic Matsubara frequency with $T$ being the temperature of the
system, ${\cal T}_\tau$ is the imaginary time-ordering operator and

$\hat{a}$ and $\hat{b}$ stand for either a differential operator
(contracted with Dirac $\gamma$-matrices) or an interaction vertex
$\Gamma^0_{s/ps}$ appearing in
Eqs.~\eqref{eq:lagrangian}--\eqref{eq:current}.

The required retarded correlation functions can be
obtained from Eq.~\eqref{eq:rings_ab} by an analytic continuation
$i\omega_n\to \omega +i\delta$.  The procedure of computation of the
bulk viscosity is slightly more involved than that of the
conductivities and shear viscosity~\cite{2008JPhG...35c5003I,
  2008EPJA...38...97A,LW14,LKW15,LangDiss,2016PhRvC..93d5205G} because
the single-loop approximation to Eq.~\eqref{eq:rings_ab} does not
cover all the relevant diagrams in the $1/N_c$ expansion. Figure
\ref{fig:loops_ab} shows the $G^0$ and $G^1$ order terms of
diagrammatic expansion for the two-point correlation function given by
Eq.~\eqref{eq:rings_ab}. To select the leading-order diagrams in the
$1/N_c$ power-counting scheme, the following rules are applied: (i) each
loop contributes a factor of $N_c$ from the trace over color space;
(ii) each coupling $G$ associated with a pair of $\Gamma^0_{s/ps}$
matrices contributes a factor of
$1/N_c$~\cite{1994PhRvC..49.3283Q,2008JPhG...35c5003I,2008EPJA...38...97A,LW14,LKW15,2016PhRvC..93d5205G,LangDiss}. Applying 
these rules we conclude that the diagrams in the first and the second lines
in Fig.~\ref{fig:loops_ab} are of order of $N_c$. The third diagram,
which is a first-order vertex correction, is of order of $N_c^0$, and,
therefore, is suppressed compared to the previous ones. Thus, the
correlation function \eqref{eq:rings_ab} in the leading
[${\cal O}(N_c^1)$] order is given by an infinite sum of bubble
diagrams, each of which consists of several single-loop
diagrams. The latter in the momentum space is given by (see the first line in Fig.~\ref{fig:loops_ab})
\bea\label{eq:ring1_ab}
\Pi_0[\hat{a},\hat{b}](\omega_n)&\equiv &-
T\sum_l\!\! \int\!\! \frac{d\vecp}{(2\pi)^3}\nonumber\\
&\times &\Tr \left[\hat{a}G(\vecp, i\omega_l+i\omega_n) \hat{b} G(\vecp, i\omega_l) \right].
\eea
Here $G(\vecp, i\omega_l)$ is the dressed Matsubara quark-antiquark
propagator, the summation goes over fermionic Matsubara frequencies
$\omega_l=\pi(2l+1)T-i\mu$, $l=0,\pm1,\ldots,$ with $\mu$ being the
chemical potential, and $\hat{a}$ and $\hat{b}$ are the momentum-space
counterparts of the same operators appearing in
Eq.~\eqref{eq:rings_ab}. The traces should be taken in Dirac, color,
and flavor space. The details of these computations and the loop
resummation are relegated to Appendix~\ref{app:A}.
 
To express the correlation functions given by Eq.~\eqref{eq:ring1_ab}
in terms of the Lorentz components of the spectral function we write
the full quark retarded/advanced Green's function as
\bea\label{eq:propagator1} 
G^{R/A}(p_0,\bm p)=\frac{1}{\slashed p-m-\Sigma^{R/A}(p_0,\bm p)},
\eea
where $m$ is the constituent quark mass, $\Sigma^{R/A}$ in (\ref{eq:propagator1}) is the quark
retarded/advanced self-energy which is written in terms of its Lorentz
components as
\bea\label{eq:selfenergy}
\Sigma^{R(A)}
=m\Sigma_s^{(*)}
-p_0\gamma_0\Sigma_0^{(*)}
+\bm p\bm\gamma\Sigma_v^{(*)}.
\eea 
By definition, the spectral function is given by
\bea\label{eq:spectralfunction}
A(p_0,\bm p)&=&-\frac{1}{2\pi i}
[G^R(p_0,\bm p)-G^A(p_0,\bm p)]\nonumber\\
&=&-\frac{1}{\pi}(mA_s+p_0\gamma_0 A_0 
-\bm p\bm\gamma A_v),
\eea 
where the scalar $A_s$, temporal $A_0$ and vector $A_v$ components
are expressed through combinations of the components (real and
imaginary) of the self-energy according to the relations~\cite{LKW15,2017arXiv170204291H}
\bea\label{eq:spectral_coeff}
A_i(p_0, p)=\frac{1}{d} 
[n_1\varrho_i -2n_2 (1+r_i) ],\quad 
d=n_1^2+4n_2^2,
\eea
with
\bea
\label{eq:N1}
n_1&=&p_0^2[(1+r_0)^2-\varrho_0^2]\nonumber\\
&-&\bm p^2[(1+r_v)^2-\varrho_v^2]
-m^2[(1+r_s)^2-\varrho_s^2],\\
\label{eq:N2}
n_2&=&p_0^2\varrho_0 (1+r_0) \nonumber\\
&-&\bm p^2\varrho_v(1+r_v)-m^2\varrho_s(1+r_s),
\eea
where we used the shorthand notations $\varrho_i = {\rm Im}\Sigma_i$ and
$r_i = {\rm Re}\Sigma_i$, $i=s,0,v$. From now on we will neglect the
irrelevant real parts of the self-energy, which lead to
momentum-dependent corrections to the constituent quark mass in
next-to-leading order ${\cal O} (N_c^{-1})$.

The bulk viscosity in terms of the components of the spectral function
is then written as 
\bea\label{eq:zeta_sum} \zeta = \zeta_0 +\zeta_1
+\zeta_2, 
\eea 
with the one-loop contribution given by
\bea\label{eq:zeta0_final} 
\zeta_0 &=&
-\frac{2N_cN_f}{9\pi^3} \int_{-\infty}^{\infty} d\varepsilon
\frac{\partial n}{\partial\varepsilon} \int_0^\Lambda dpp^2 \Big[2(ax+by+cz)^2 \nonumber\\
&&-(x^2-y^2+z^2)(a^2-b^2+c^2)\Big],
\eea
where $N_c=3$ and $N_f=2$ are the color and flavor numbers,
respectively, and
\be\label{eq:xyz}
x=3(1+\gamma)m_0,\quad y=3(\delta-\varepsilon), \quad z=(2+3\gamma)p,
\ee
\be
\label{eq:abc}
a=mA_s,\quad b=\varepsilon A_0,\quad c=pA_v.
\ee
In Eq.~\eqref{eq:zeta0_final} we introduced a regularizing 3-momentum
ultraviolet cutoff $\Lambda$; below we adopt the value
$\Lambda= 0.65$ GeV.  The quark distribution function is given by
\bea\label{eq:fermi}
n(\varepsilon)=\frac{1}{e^{\beta(\varepsilon-\mu)}+1},
\eea
with $\beta=T^{-1}$ being the inverse temperature.
 The following two
contributions in Eq.~\eqref{eq:zeta_sum} are given by
\bea\label{eq:zeta_12_final}
\zeta_1 =2(\bar G\bar R )I_1,\qquad \zeta_2 =  (\bar G\bar R )^2I_2,
\eea
where the renormalized coupling $\bar G$ arises through resummation of
geometrical series as
\bea\label{eq:G_bar0}
\bar G = \frac{G}{1-R_0G},
\eea
with the polarization loop
\bea\label{eq:R_0}
R_0&=&-\frac{2N_cN_f}{\pi^4}
\int_{-\infty}^{\infty} d\varepsilon 
\int_{-\infty}^{\infty} d\varepsilon' 
\frac{n(\varepsilon)-n(\varepsilon')}
{\varepsilon-\varepsilon'}
\nonumber\\
&&\times \int_0^\Lambda dpp^2 (aa'+bb'-cc').
\eea
Finally, the three functions appearing in Eq.~\eqref{eq:zeta_12_final} are
given by
\bea 
\label{eq:I1}
I_1&=&-\frac{2N_cN_f}{3\pi^3}
\int_{-\infty}^{\infty} d\varepsilon 
\frac{\partial n}{\partial\varepsilon}
\int_0^\Lambda dpp^2\nonumber\\
&& \times
\Big[x(a^2+b^2-c^2)+2a(by+cz)\Big],\\
\label{eq:I2}
I_2&=&-\frac{2N_cN_f}
{\pi^3 }\int_{-\infty}^{\infty} 
d\varepsilon \frac{\partial n}{\partial\varepsilon} \int_0^\Lambda dpp^2 
(a^2+b^2-c^2),\\
\label{eq:R_bar}
\bar R &=&-\frac{2N_cN_f}{3\pi^4}
\int_{-\infty}^{\infty} d\varepsilon 
\int_{-\infty}^{\infty} d\varepsilon' 
\int_0^\Lambda dp {p^2}
\frac{1}{\varepsilon-\varepsilon'}\nonumber\\
&\times&\Big\{[n(\varepsilon)-n(\varepsilon')]
\big[x(aa'+ bb'-cc')+z(a'c+ac')\big]\nonumber\\
&&+\Big[yn(\varepsilon)-y'n(\varepsilon')+\frac{3}{2}(\varepsilon-\varepsilon')\Big]
(a'b+ ab')\Big\}.
\eea 
Here the functions $a',b',c',y'$ are obtained from $a,b,c,y$ defined
in Eqs.~\eqref{eq:xyz} and \eqref{eq:abc} by substitution
$\varepsilon\to\varepsilon'$.  Equations
\eqref{eq:zeta_sum}--\eqref{eq:R_bar} express the bulk viscosity of the
quark plasma in terms of the components of its spectral function.

It is remarkable that the multiloop
contributions do not vanish if the chiral symmetry is explicitly broken. However, 
in the chiral limit $m_0=0$
they vanish trivially, since quarks become massless
 above the critical temperature $T_c$ (see the next section). Indeed, from Eqs.~\eqref{eq:xyz} and \eqref{eq:abc} we find $x=0$ and
$a, a'\propto m=0$ in this case.
 Consequently, it follows from
Eqs.~\eqref{eq:R_bar} and \eqref{eq:zeta_12_final} that
$\zeta_{1,2}=0$. Therefore, the bulk viscosity in this case is given 
by the single-loop contribution $\zeta_0$, which remains finite also
in the chiral limit; see also the discussion in 
Sec.~\ref{sec:chiral}.

\section{Phase diagram and spectral functions}
\label{sec:QSpectralFunctions}

Here we specify the structure of the phase diagram of strongly
interacting quark matter and review the processes that lead to the
dispersive effects (imaginary parts of the self-energy of quarks and
antiquarks) within the region of $\mu$-$T$ plain. Our discussion is
based on the two-flavor NJL model described by the Lagrangian
\eqref{eq:lagrangian}.

\begin{figure}[tbh] 
\begin{center}
\includegraphics[height=1.8cm]{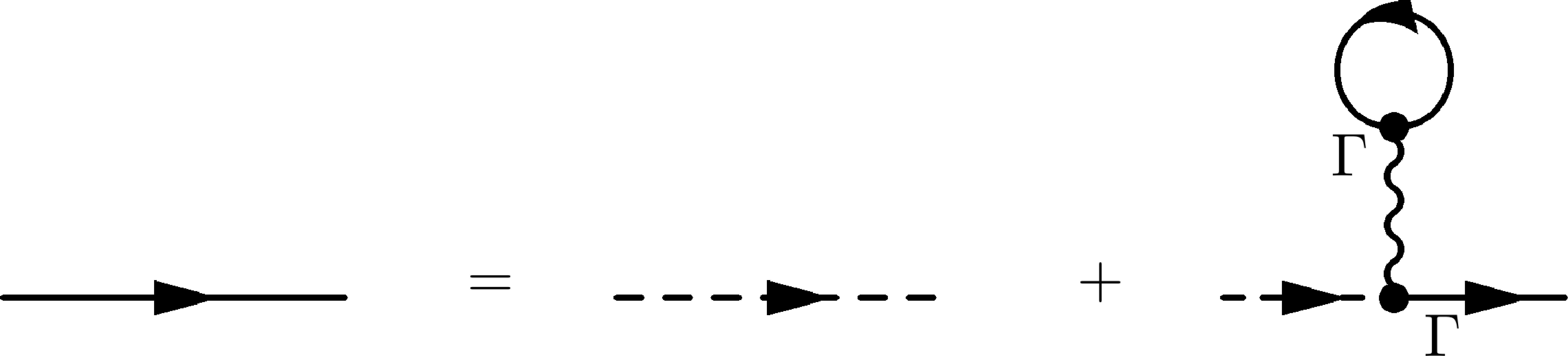}
\caption{
The Dyson-Schwinger equation for the
  constituent quark mass.  The dashed and solid lines stand for the
  bare and dressed propagators, respectively, and the vertex
  $\Gamma=1$. The wavy line represents the interaction.}
\label{fig:gap_eq} 
\end{center}
\end{figure}

Within the NJL model the nonzero temperature and density constituent
quark mass is determined to leading order in the $1/N_c$ expansion from a
Dyson-Schwinger equation, where the self-energy is taken in the
Hartree approximation in terms of a tadpole diagram (so called quark
condensate), see Fig.~\ref{fig:gap_eq}.  From Fig.~\ref{fig:gap_eq} we
obtain the following equation for the constituent quark mass
\bea\label{eq:mass} m=m_0-G\langle\bar{\psi}\psi\rangle,  
\eea
where the quark condensate is given by 
\bea\label{eq:gap3}
\langle\bar{\psi}\psi\rangle= -\frac{mN_cN_f}{\pi^2}\int_0^\Lambda
dp\frac{p^2}{E_p}[1-n^+(E_p)-n^-(E_p)], \nonumber\\
\eea 
with quark/antiquark thermal distributions
$n^\pm(E)=[e^{\beta(E\mp\mu)}+1]^{-1}$.

\begin{figure}[bth] 
\begin{center}
\includegraphics[width=8.0cm,keepaspectratio]{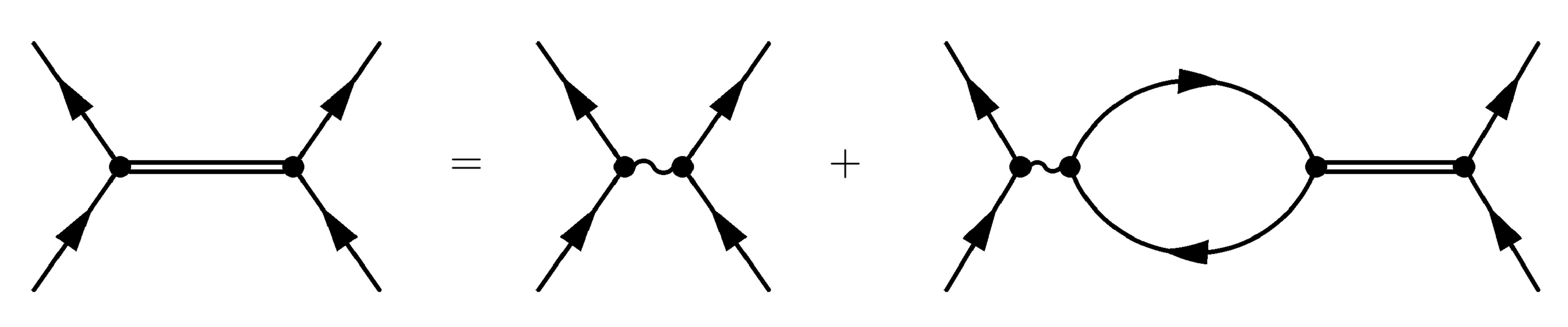}
\caption{The Bethe-Salpeter equation for mesons: the double lines stand
  for the dressed meson propagators. The remaining diagrammatic
  elements are as in Fig.~\ref{fig:gap_eq}, except that the vertex assumes
  the values $\Gamma^0_{s}=1$ for $\sigma$ meson and $\Gamma^0_{ps}=i\bm\tau\gamma_5$
  for pions. }
\label{fig:BS_eq} 
\end{center}
\end{figure}

 The propagators of $\pi$ and $\sigma$ mesons are
found from the Bethe-Salpeter equation illustrated in
Fig.~\ref{fig:BS_eq}, which resums contributions from quark-antiquark polarization insertions.


Once the in-medium propagator of mesons is found, their
masses are then determined from the propagator poles in real
spacetime for $\bm p=0$; for details see~\cite{2017arXiv170204291H}
and references therein.

\begin{figure}[t] 
\begin{center}
\includegraphics[width=7.5cm,keepaspectratio]{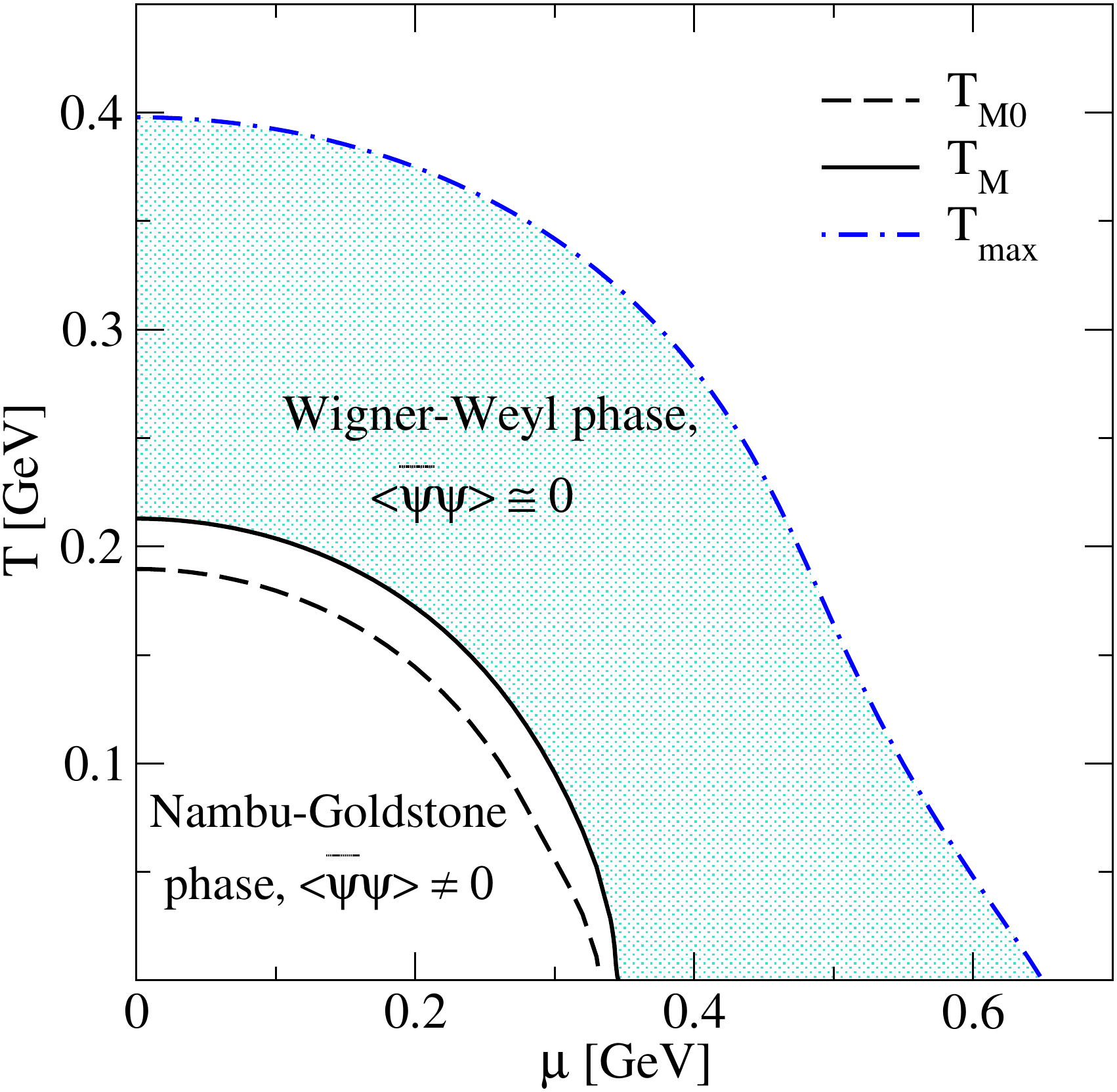}
\caption{ The shaded area shows the region of the phase diagram of
  strongly interacting quark matter where our computations are applicable.
  The area is bound by the Mott temperature $T_{\rm M}$ in the case of
  broken chiral symmetry or $T_{\rm M 0}\equiv T_c$ in the case where
  chiral symmetry is intact and by the maximal temperature
  $T_{\rm max}$ above which no meson modes are found. }
\label{fig:mott_temp1} 
\end{center}
\end{figure}

The region of the $\mu$-$T$ plain where our model is applicable is
shown in Fig.~\ref{fig:mott_temp1} by the shaded area. Its outer
boundary is given by the maximal temperature $T_{\rm max}$ above which
no solutions for meson masses can be found. More precisely, mesonic
modes do not exist for $T\ge T_{\rm max}$ within our zero-momentum
pole approximation.  In the case of $T = 0$ the transition line ends at
the maximal value of the chemical potential $\mu_{\rm max}=\Lambda$
where the meson mass $m_M = 2\Lambda$. The inner boundary of the
$\mu$-$T$ region corresponds to the so-called Mott temperature
$T_{\rm M}$ at which the condition $m_\pi=2m$ is fulfilled. The Mott
temperatures for the cases where chiral symmetry is intact ($m_0=0$)
and chiral symmetry is explicitly broken ($m_0\neq 0$) differ only
slightly, see Fig.~\ref{fig:mott_temp1}. The dispersive effects of
interest which correspond to meson decays $\pi,\sigma\to q+\bar{q}$ and the inverse
processes are allowed kinematically above $T_{\rm M}$ for a given
$\mu$. Note that in the chiral limit $m_0=0$
the Mott temperature coincides with the critical
temperature $T_c$ of the chiral phase transition, above which we have $\langle\bar{\psi}\psi\rangle=0$ and $m=0$.
\begin{figure*}[!] 
\begin{center}
\includegraphics[width=14.0cm,keepaspectratio]{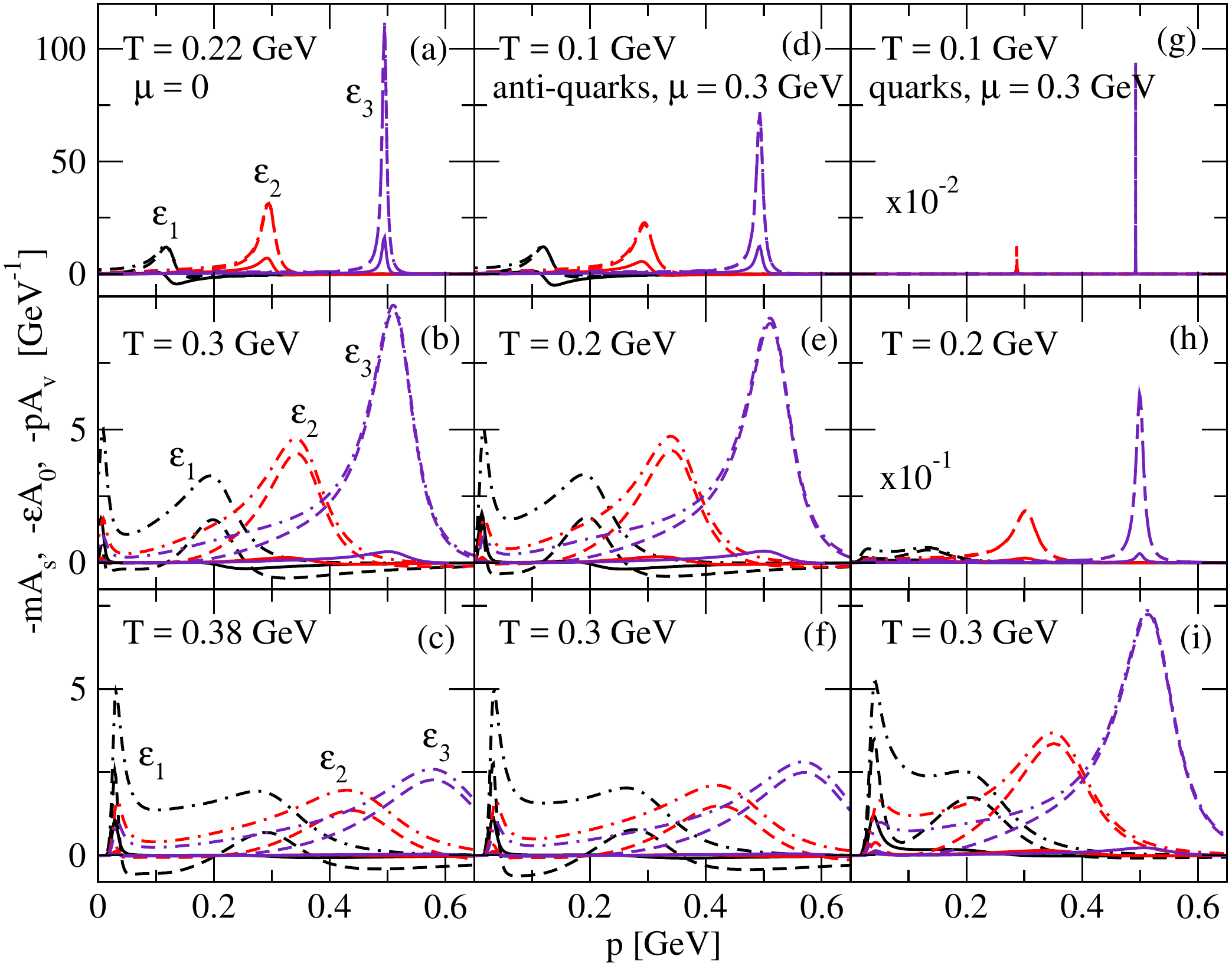}
\caption{ The Lorentz components of the quark and antiquark spectral
  functions $-mA_s$ (solid line), $-\varepsilon A_0$ (dash-dotted
  line) and $-pA_v$ (dashed line) as functions of momentum for fixed
  values of energy.  The panels (a)--(c) correspond to $\mu=0$,
  (d)--(f) -- to antiquarks with $\mu=0.3$ GeV, and (g)--(i) -- to quarks with
  $\mu=0.3$ GeV. The spectral functions are evaluated at three
  energies $\varepsilon_1 = 0.1$, $\varepsilon_2 = 0.3$, and
  $\varepsilon_3 = 0.5$ GeV.  }
\label{fig:spectral} 
\end{center}
\end{figure*}

The quark self-energy corresponding to the meson decays into two
quarks and the inverse processes within the regime of interest is given in Matsubara space by~\cite{LKW15,2017arXiv170204291H}
\bea
  \Sigma^M(\bm p,\omega_n) &=&
 \begin{minipage}{3cm}
\includegraphics[width=1.\textwidth]{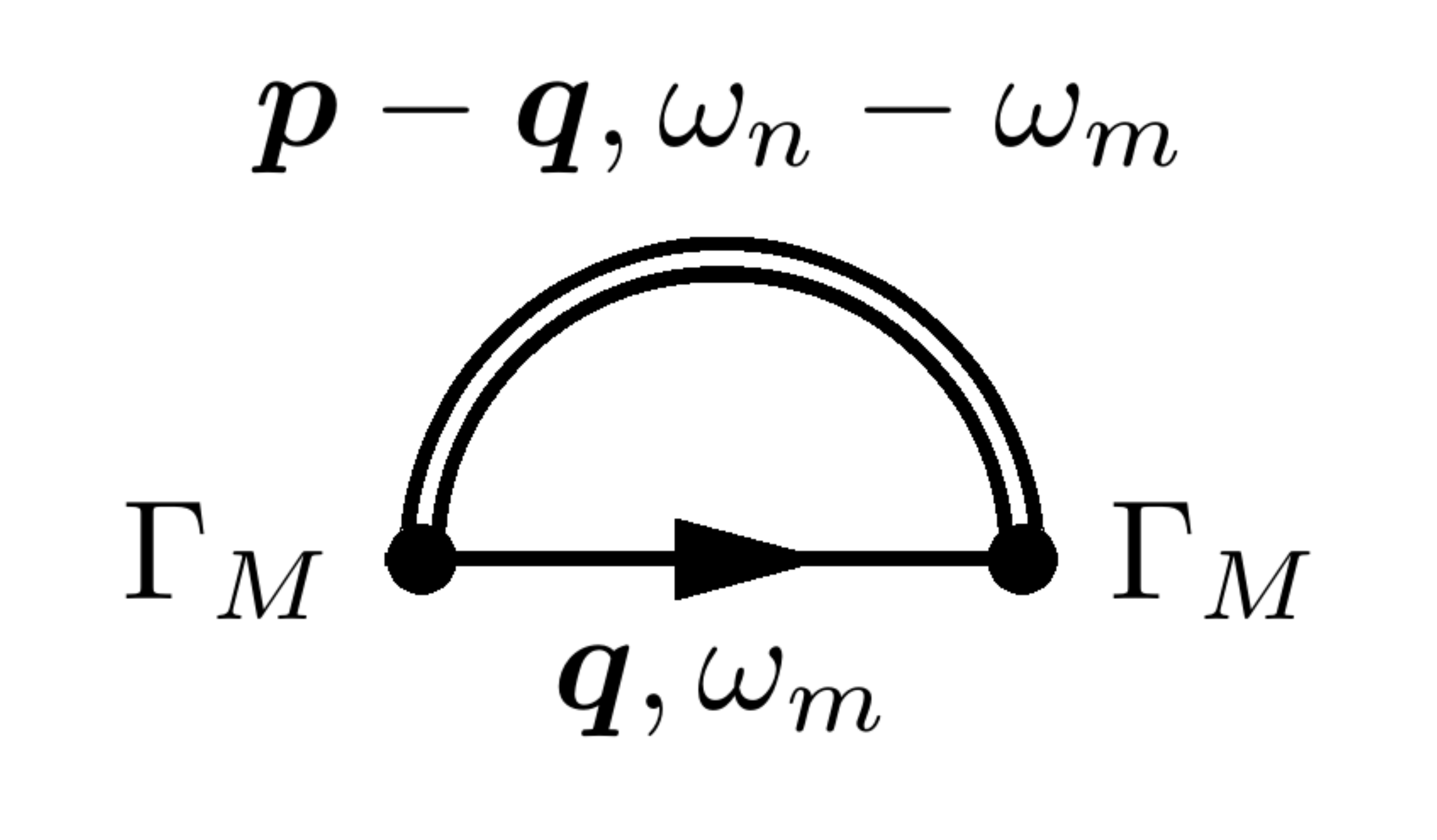} 
\end{minipage}
\nonumber\\
&&\hspace{-2.3cm} =T \sum_m \int \frac{d{\bm q}}{(2\pi)^3} 
\left[\Gamma_M S(\bm q, \omega_m) 
\Gamma_M D_M(\bm p-\bm
  q,\omega_n-\omega_m) \right],\nonumber\\
\label{eq:self}
\eea
where $S(\bm q, \omega_m)$ is the quark propagator with constituent quark mass, the index $M=\pi, \sigma$ stands for the $\pi$ and $\sigma$
mesons and the vertices are given by $\Gamma_\sigma=1$ and
$\Gamma_\pi=i\gamma_5\bm\tau$. The Lorentz decomposition of the
Matsubara self-energy, which is analogous to \eqref{eq:selfenergy},
is given by
\bea\label{eq:self1}
\Sigma^M(\bm p,\omega_n)=
P_Mm\Sigma^M_s+i\omega_n\gamma_0\Sigma^M_0-\bm p \cdot \bm \gamma\Sigma^M_v, 
\eea
with $P_\sigma=1$, $P_\pi=-1$. The computation of the components of
this decomposition gives~\cite{LKW15,2017arXiv170204291H}
\bea\label{eq:self_sv1}
\Sigma^M_{s,v} &=&
g^2_M\int \frac{d\bm q}{(2\pi)^3}
\frac{\mathscr{Q}_{s,v}}{4E_qE_M} \nonumber\\
&\times&\left[\frac{i\omega_n \mathscr{C}_3-2E_+ \mathscr{C}_1}
{E_+^2+\omega_n^2}-
\frac{i\omega_n \mathscr{C}_3+2E_- \mathscr{C}_2}
{E_-^2+\omega_n^2}\right],\\
\label{eq:self_01}
\Sigma^M_0 &=&
g^2_M\int \frac{d\bm q}{(2\pi)^3} 
\frac{\mathscr{Q}_{0}}{4 E_qE_M}\nonumber\\
&\times&\left[\frac{2i\omega_n \mathscr{C}_1-E_+\mathscr{C}_3}{E_+^2+\omega_n^2}+
\frac{2i\omega_n \mathscr{C}_2+E_-\mathscr{C}_3}
{E_-^2+\omega_n^2}\right],
\eea
where $g_M$ is the quark-meson coupling constant and we defined shorthand notations
\bea
&&\mathscr{C}_1=1+n_B(E_M)-\frac{1}{2}[n^+(E_q)+n^-(E_q)],\nonumber\\
\label{eq:Z_123}
&&\mathscr{C}_2=n_B(E_M)+\frac{1}{2}[n^+(E_q)+n^-(E_q)],\\
&&\mathscr{C}_3=n^+(E_q)-n^-(E_q),\nonumber
\eea
and 
\bea\label{eq:f_sv}
\mathscr{Q}_{s}=1,\quad
\mathscr{Q}_{v}=\frac{\bm q\cdot\bm p}{p^2},\quad
\mathscr{Q}_{0}=-\frac{E_q}{i\omega_n},
\eea
with $E_\pm=E_q\pm E_{q+p}$, $E_M=E+E_p$, and $E_p =\sqrt{p^2+m^2}$.
The distribution functions of quarks and antiquarks are defined as
$n^\pm(E)=[e^{\beta(E \mp \mu)}+1]^{-1}$, and
$n_B(E)=(e^{\beta E}-1)^{-1}$ is the Bose distribution function for mesons at
zero chemical potential.
The retarded self-energy is now obtained by analytical continuation
$i\omega_n\to p_0+i\varepsilon$ and has the same Lorentz structure as
its Matsubara counterpart.  
For the imaginary part of the on-shell quark and antiquark
self-energies
($\varrho \equiv {\rm Im}\Sigma$) one finds~\cite{LKW15,2017arXiv170204291H}
\bea\label{eq:im_self}
\varrho_{j}^M(p)\Big\vert_{p_0=E_p}\!\!\! &=&
\frac{g^2_M}{16\pi p}
\int_{E_{\rm min}}^{E_{\rm max}}\!\!\! d E
\nonumber\\
&&\hspace{1.cm}\times\mathscr{T}_{j}[n_B(E_M)+n^-(E)],\quad\\
\label{eq:im_self_anti}
\varrho^M_{j}(p)\Big\vert_{p_0=-E_p}\!\!\!\!&=&-
\frac{g^2_M}{16\pi p}
\int_{E_{\rm min}}^{E_{\rm max}}\!\!\!\!d E \nonumber\\
&&\hspace{1.cm}\times
\mathscr{T}_{j}[n_B(E_M)+n^+(E)],\quad
\eea
where $j=s,0,v,$  $E_M=E+E_p$, and
\bea\label{eq:f_sv}
\mathscr{T}_{s}=1,\quad
\mathscr{T}_{v}=\frac{m_M^2-2m^2-2EE_p}{2p^2},\quad
\mathscr{T}_{0}=-\frac{E}{E_p}.
\eea
The integration limits are defined as 
\bea\label{eq:E_min_max}
 E_{{\rm min},{\rm max}}&=&
\frac{1}{2m^2}\left[(m_M^2-2m^2)  p_0 \right. \nonumber \\
&  & \hspace*{1cm} \left. \pm pm_M\sqrt{m_M^2-4m^2}\right],
\eea  
and in the chiral limit $m=0$
\bea\label{eq:E_min_max_chiral}
 E_{\rm min}=\frac{m_M^2}{4p},\quad
 E_{\rm max}\to \infty.
\eea
We stress here that
Eqs.~\eqref{eq:im_self}--\eqref{eq:E_min_max_chiral} are applicable
only above the Mott (critical) temperature, where the condition
$m_M\ge 2m$ is fulfilled. Finally, the full quark-antiquark
self-energy in on-shell approximation is written as
\bea\label{eq:im_self_onshell}
\varrho_j(p_0,p) =
\theta (p_0)\varrho^+_j(p)+
\theta (-p_0)\varrho^-_j(p),
\eea
with $\varrho^\pm_j(p)=\varrho_j(p_0=\pm E_p,p)$.
From Eqs.~\eqref{eq:im_self} and \eqref{eq:im_self_anti} it follows that
$\varrho^+$ and $\varrho^-$ obey the relation 
$\varrho^-_j(\mu, p)=-\varrho^+_j(-\mu, p)$, and, consequently, 
\bea\label{eq:relation_rho}
\varrho_j(\mu, -p_0, p) &=&-\varrho_j(-\mu, p_0, p).
\eea
The contribution of the mesons to the net quark/antiquark self-energy
is summed as follows
\bea\label{eq:selfsum2} 
\Sigma_s= \Sigma^\sigma_s-3\Sigma^\pi_s,\quad
\Sigma_{0/v} =-\Sigma^\sigma_{0/v} -3\Sigma^\pi_{0/v}.
\eea
In the final step the spectral functions of quarks and antiquarks are
constructed according to the relations \eqref{eq:spectral_coeff}--\eqref{eq:N2}, where we neglect the real parts which
are higher order in the power counting scheme.
The numerical results for the components of the spectral function
are shown in Fig.~\ref{fig:spectral} and will be used below in the 
computations of the bulk viscosity.

The key features of spectral functions which are shown for three
values of the quark (off-shell) energy
($\varepsilon_1 = 0.1$,
$\varepsilon_2 = 0.3$ and $\varepsilon_3 = 0.5$ GeV) are as follows:
(a) the spectral functions display a peak at the values of momenta
$p\simeq \varepsilon$, which can be anticipated from
Eqs.~\eqref{eq:spectral_coeff}--\eqref{eq:N2} and is a consequence of the fact that the
denominator $d$ attains its minimum roughly at $p\simeq p_0$
($p_0\equiv \varepsilon$); (b) the heights of the peaks universally
increase with the (off-shell) energies of the quarks; (c) with
increasing temperature the dispersive effects become more pronounced,
consequently the quasiparticle peaks become broader and the Lorentzian
shape of the spectral functions develops in a more complex structure;
(d) the main contribution to the spectral function comes from the
temporal and vector components, which contribute comparable amounts,
whereas the scalar component is small; and (e) the quasiparticle peaks are
sharper for quarks rather than for antiquarks for the same values of
temperature and chemical potential. Note also that while the Lorentz
components of the spectral function may change the sign, the width of
the quasiparticles, which is a combination of these, remains positive,
which guarantees the overall stability of the system~\cite{LangDiss}.

\section{Numerical results for bulk viscosity}
\label{sec:results}

We start our analysis with an examination of the influence of various
factors entering the expressions for bulk viscosities $\zeta_0$,
$\zeta_1$ and $\zeta_2$.  Readers interested only in the results on
the bulk viscosity can skip to the following subsection.

\subsection{Preliminaries}

The behavior of the two-dimensional integrals determining $\zeta_0$,
$I_1$ and $I_2$ through Eqs.~\eqref{eq:zeta0_final}, \eqref{eq:I1} and
\eqref{eq:I2} is as follows.  For a given value of $\varepsilon$ the
inner integrands are peaked at $p\simeq |\varepsilon|$, as implied by
the shape of the spectral functions. The heights of the peaks rapidly
increase with the value of $|\varepsilon|$. As a consequence, the
(inner) momentum integrals are increasing functions of $|\varepsilon|$
for $\vert\varepsilon\vert\leq\Lambda$.  For energies larger than
$\Lambda$ the peaks are outside of the momentum-integration range
(because of the momentum cutoff $p\leq\Lambda$), and the momentum
integral rapidly decreases with $|\varepsilon|$.  It vanishes
asymptotically in the limit $\varepsilon\to \pm\infty$ for $I_1$ and
$I_2$, but tends to a constant value for $\zeta_0$. This asymptotic
behavior is easily seen from Eq.~\eqref{eq:zeta0_final}. Its inner
integrand can be roughly estimated as
$\propto p^2\Big[2(\varepsilon^2 A_0 -p^2A_v)^2- (\varepsilon^2-p^2)
(\varepsilon^2A_0^2-p^2A_v^2)\Big]=p^2\Big[(\varepsilon^2 A_0
-p^2A_v)^2+ \varepsilon^2p^2(A_0-A_v)^2\Big]$,
where we approximated $\gamma \simeq 1/3$ and $\delta\simeq 0$
(see Appendix \ref{app:B}) and neglected the scalar component of the
spectral function, which is small compared to the vector and temporal
components. If $|\varepsilon|\gg p$, we can approximate
Eqs.~\eqref{eq:N1} and \eqref{eq:N2} as
$n_1=\varepsilon^2(1-\varrho_0^2)$, $n_2=\varepsilon^2\varrho_0$. The
dominant term in the integrand in this case is
$\propto p^2\varepsilon^4 A_0^2=p^2\varepsilon^4(\varrho_0
n_1-2n_2)^2/(n_1^2+4n_2^2)^2=p^2\varrho_0^2/(1+\varrho_0^2)^2$,
which does not depend on $\varepsilon$ in the on-shell approximation
to the self-energy.  As a result, the momentum integral tends to a
constant value for $|\varepsilon|\ge \Lambda$.  The outer integrals of
Eqs.~\eqref{eq:zeta0_final}, \eqref{eq:I1} and \eqref{eq:I2} contain
the Fermi factor $\partial n(\varepsilon)/\partial\varepsilon$ which
at low temperatures is strongly peaked at the energy
$\varepsilon = \mu$.  At high temperatures it transforms into a
bell-shaped broad structure which samples energies away from $\mu$. We
have verified numerically that it is sufficient to integrate up to the
energy $|\varepsilon|\le 2$~GeV.  Next we note that the outer integral
samples the contribution of antiquarks from the range $(-\infty,0)$
and that of quarks from the range $(0,+\infty)$, and we are in a position to
examine these two contributions separately. We
find that when $\mu = 0$, the integrands of
Eqs.~\eqref{eq:zeta0_final}, \eqref{eq:I1}, and \eqref{eq:I2} are even
functions of $\varepsilon$, \ie, the quark and antiquark
contributions are the same. At nonzero chemical potentials the
quark-antiquark symmetry is broken and the contributions from quarks
and antiquarks differ. While the contributions of quarks and
antiquarks to the (inner) momentum integrands are comparable at
nonzero $\mu$, the factor $\partial n/\partial\varepsilon$ in the
outer energy integration makes the quark contribution dominant.

\begin{figure}[tbh] 
\begin{center}
\includegraphics[width=8.cm,keepaspectratio]{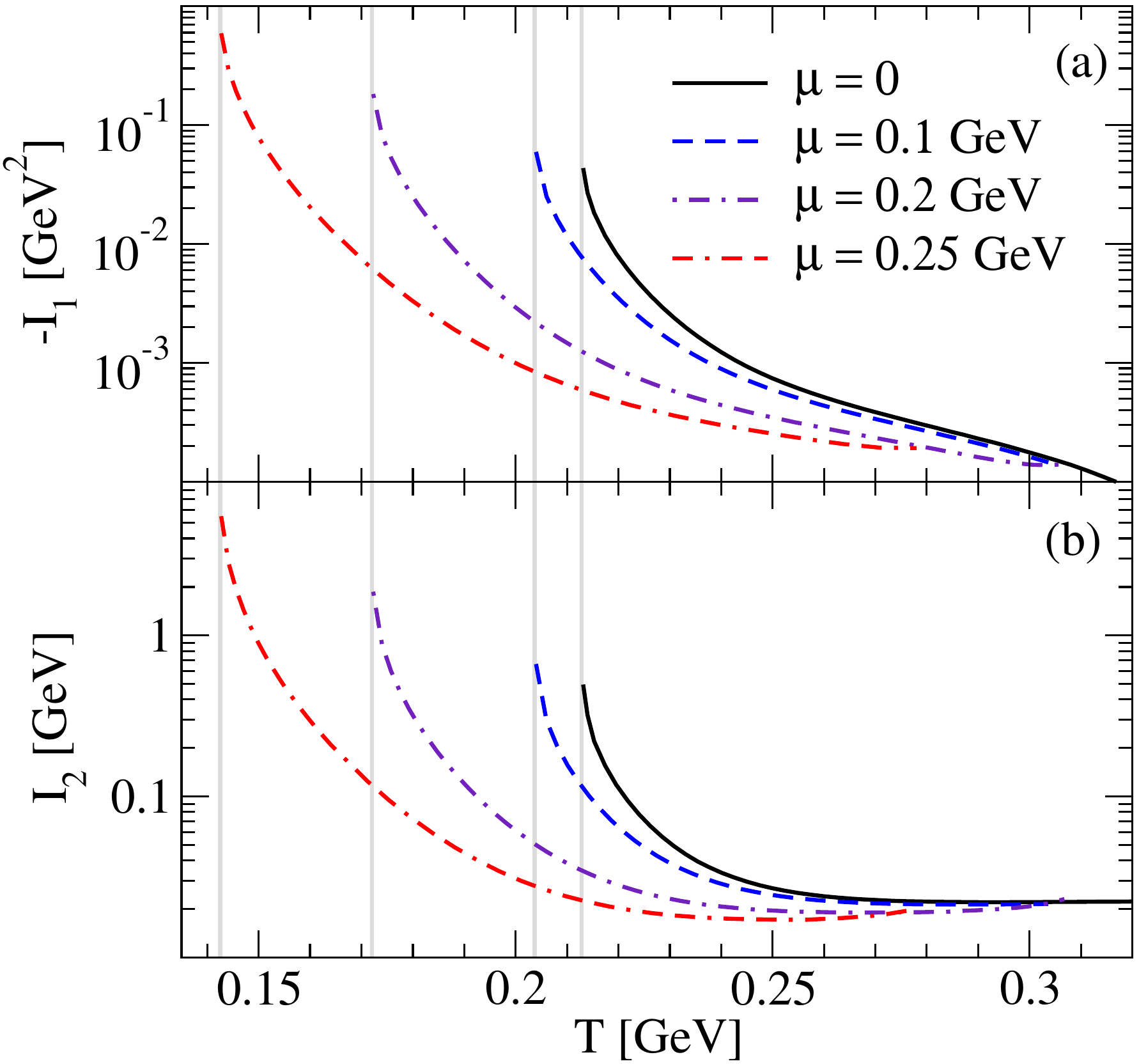}
\caption{Dependence of the integrals $I_1$ and $I_2$ on the
  temperature for several values of the chemical potential. The
  vertical lines show the Mott temperature at the given value of
  $\mu$. }
\label{fig:I12_T} 
\end{center}
\end{figure}
\begin{figure}[!] 
\begin{center}
\includegraphics[width=8.cm,keepaspectratio]{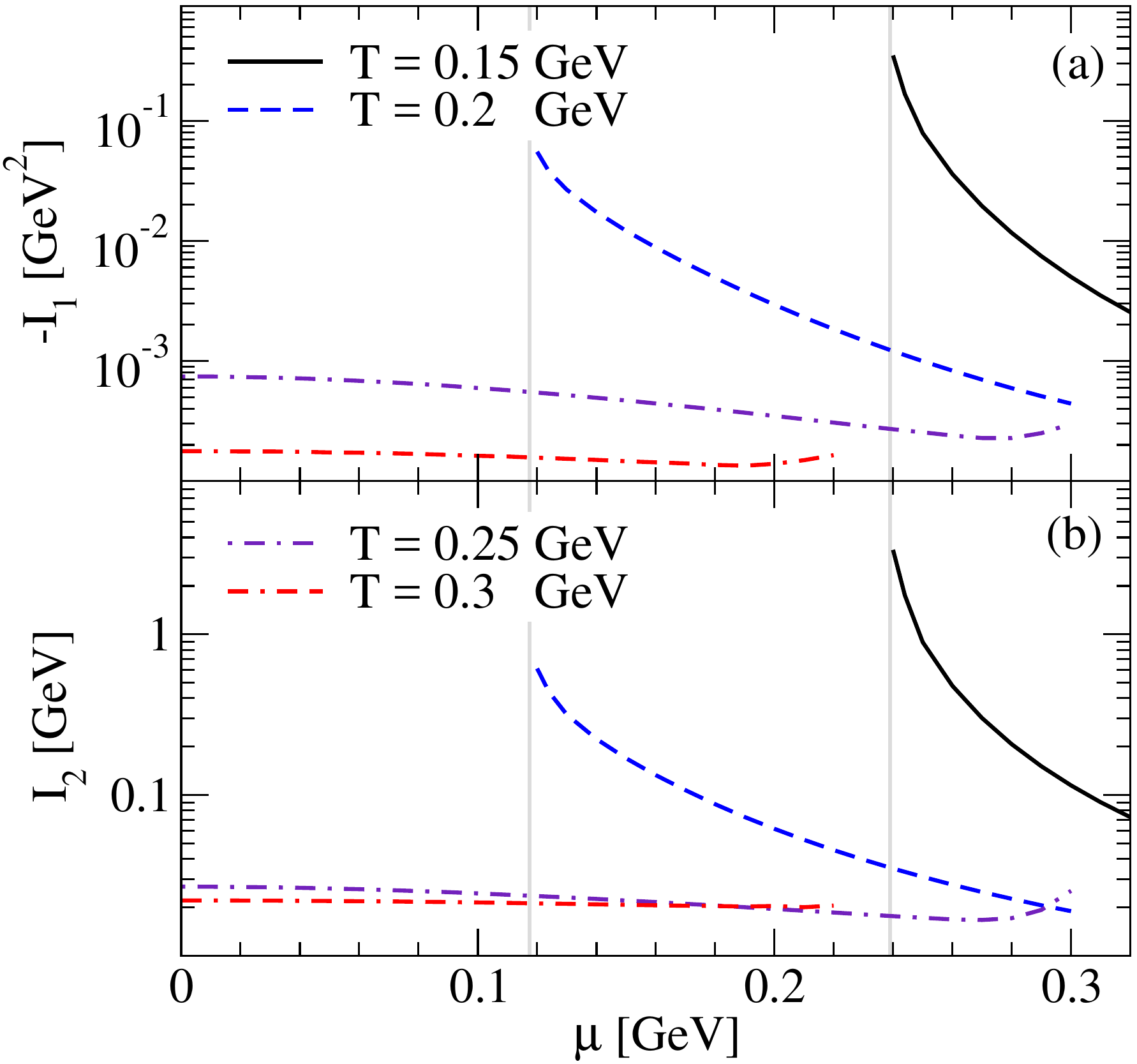}
\caption{ Dependence of the integrals $I_1$ and $I_2$ on the chemical
  potential for several values of the temperature.  The vertical lines
  show the values of the chemical potential where the temperature
  approaches the Mott temperature. }
\label{fig:I12_mu} 
\end{center}
\end{figure}

The dependence of the integrals $I_1$ and $I_2$ on temperature and
chemical potential is shown in Figs.~\ref{fig:I12_T} and
\ref{fig:I12_mu}.  Both are rapidly decreasing functions of temperature
(at fixed chemical potential) or chemical potential (at fixed
temperature) in the regime close to the Mott line.  The observed
decrease is the result of broadening of the spectral functions with
the temperature, which physically corresponds to stronger dispersive
effects and, therefore, smaller values of transport coefficients.
Note that in the vicinity of the Mott temperature these quantities
become very large because the widths of the spectral functions
originating from the imaginary parts of the self-energies vanish for
pions and are very small for $\sigma$-mesons. This is partly due to
the on-shell approximation to the self-energies. Including off-shell
contribution to the self-energies improves the asymptotics close to
$T_M$, however it is unimportant at temperatures already slightly
above the Mott temperature, where the transport coefficients are
described by on-shell kinematics quite well~\cite{LKW15}. In the whole
temperature-density range considered $I_1$ is always negative, while
$I_2$ is always positive.
$-I_1$ is always a decreasing function of
the temperature, whereas $I_2$ tends to a constant value at high
temperatures for small chemical potentials, but shows a slight minimum
at higher chemical potentials.

\begin{figure}[t] 
\begin{center}
\includegraphics[width=8.cm,keepaspectratio]{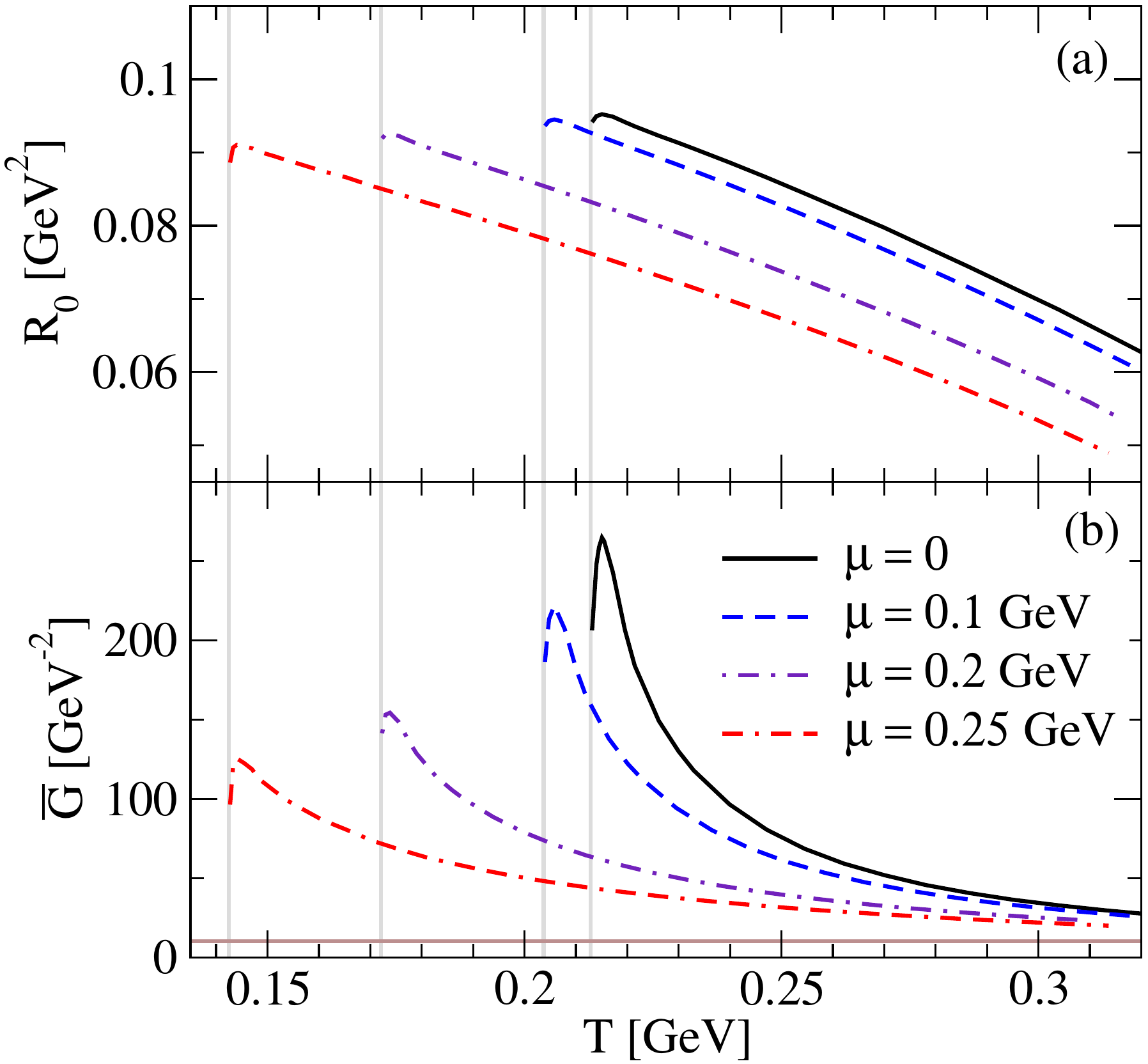}
\caption{ Integral $R_0$ (a) and the renormalized coupling $\bar G$ (b) as functions of the temperature for various values of the chemical potential. The value of the bare coupling constant $G$ is shown by solid horizontal line. } 
\label{fig:int_R0_G0_temp} 
\end{center}
\end{figure}
\begin{figure}[!] 
\begin{center}
\includegraphics[width=8.cm,keepaspectratio]{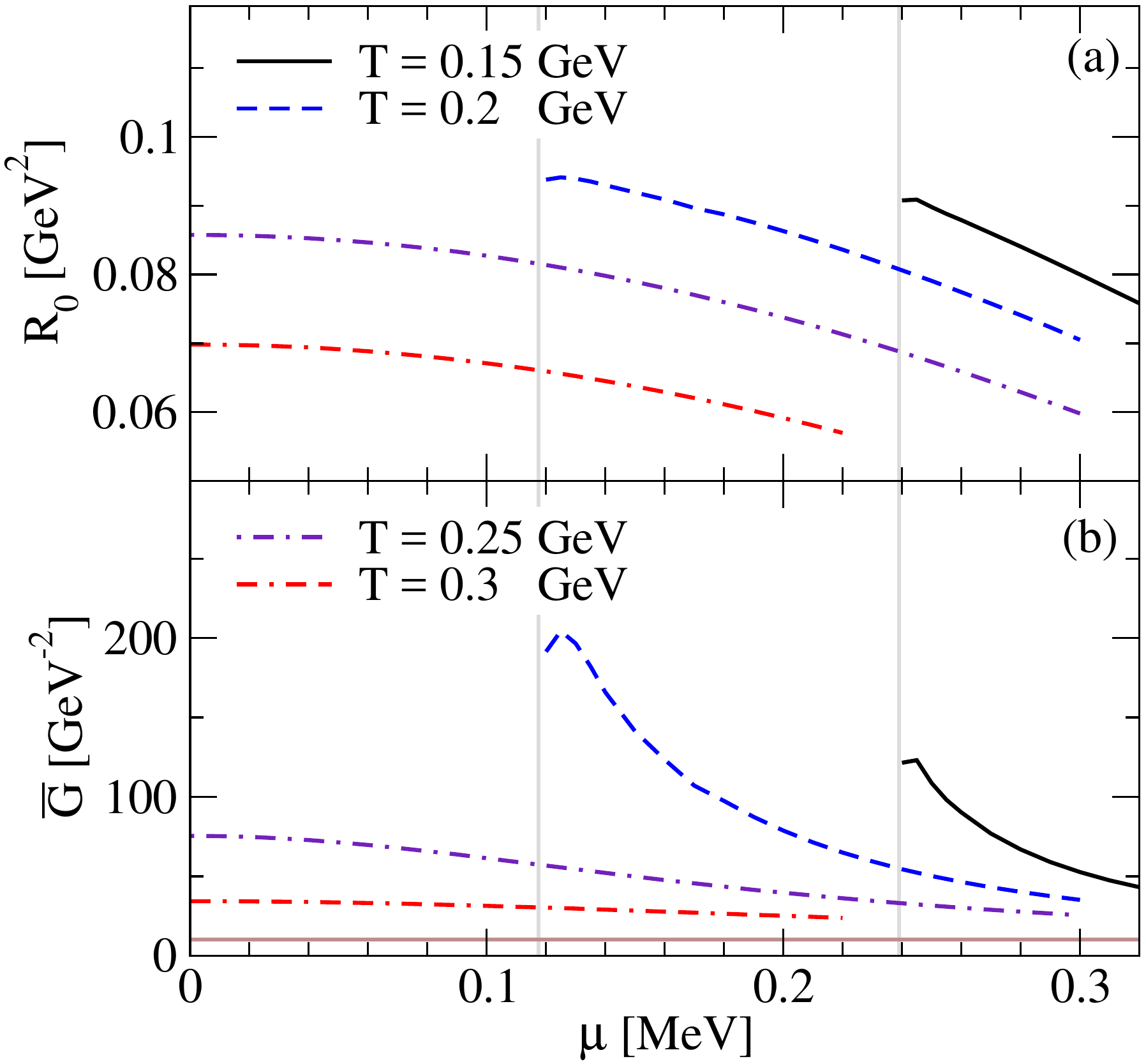}
\caption{ Integral $R_0$ (a) and the renormalized coupling $\bar G$ (b) as functions of the chemical potential
for various values of the temperature. The value of the bare coupling constant $G$ is shown by solid horizontal line. } 
\label{fig:int_R0_G0_mu} 
\end{center}
\end{figure}

Next we turn to the discussion of three-dimensional integrals $R_0$ and
$\bar R$, given by Eqs.~\eqref{eq:R_0} and \eqref{eq:R_bar}. A new
feature that appears in these expression is the convolution of two
spectral functions.  As a result, the integrands of $R_0$ and $X$ have
sharp peaks at $p\simeq |\varepsilon|$ if
$|\varepsilon|\simeq|\varepsilon'|$, and they transform into broad
structures with two smaller maxima located at $p\simeq |\varepsilon|$
and $p\simeq |\varepsilon'|$ when
$|\varepsilon|\neq|\varepsilon'|$. Therefore, the main contribution to
the integrals arises from the domain where
$p\simeq |\varepsilon|\simeq |\varepsilon'|$. Because the integration
range covers both positive and negative values of $\varepsilon$ there
are two posibilities $\varepsilon' = \pm\varepsilon$ for maximum to
arise.  In the case of $R_0$ integral only the minus sign is realized.
Indeed, because the temporal and vector components of the spectral
function have the same order of magnitude, the inner integrand of
$R_0$ can be roughly estimated as
$aa'+bb'-cc'\simeq \varepsilon\varepsilon'A_0A_0'-p^2A_vA_v'\simeq
(\varepsilon\varepsilon'-p^2)A_0A_0'$,
see Eqs.~\eqref{eq:R_0}. Therefore, the peaks around
$p\simeq\varepsilon\simeq\varepsilon'$ originating from temporal and
vector components almost cancel each other, and the momentum integral
is mainly concentrated around $\varepsilon'\simeq-\varepsilon$. The
integral $\bar R$ contains additional terms which support also a peak
at $\varepsilon'\simeq\varepsilon$ and, consequently, the momentum
integrand obtains contributions at two locations. In both cases of
$R_0$ and $\bar R$, the height of the peaks rapidly increases with the
increase of $|\varepsilon|$ as long as $|\varepsilon|\le \Lambda$ and
becomes negligible for higher values of $|\varepsilon|$. The
integration over $\varepsilon'$ contains also the factor
$[n(\varepsilon)-n(\varepsilon')]/(\varepsilon-\varepsilon')$ which at
low temperatures is strongly peaked at the energies
$\varepsilon = \varepsilon'=\mu$. At high temperatures it transforms
into a bell-shaped broad structure (without change of the location of
the maximum) and samples energies far away from $\mu$. It decreases
faster at high energies in the case when $\varepsilon-\mu$ and
$\varepsilon'-\mu$ have the same sign.  The integrand of $\bar R$
contains an additional combination of Fermi functions
$[\varepsilon n(\varepsilon)-\varepsilon'
n(\varepsilon')]/(\varepsilon-\varepsilon')-1/2$,
which tends to the finite limits $-1/2$ and $1/2$, when
$\varepsilon,\varepsilon'\to +\infty$ and
$\varepsilon,\varepsilon'\to -\infty$, respectively.

\begin{figure}[t] 
\begin{center}
\includegraphics[width=8.cm,keepaspectratio]{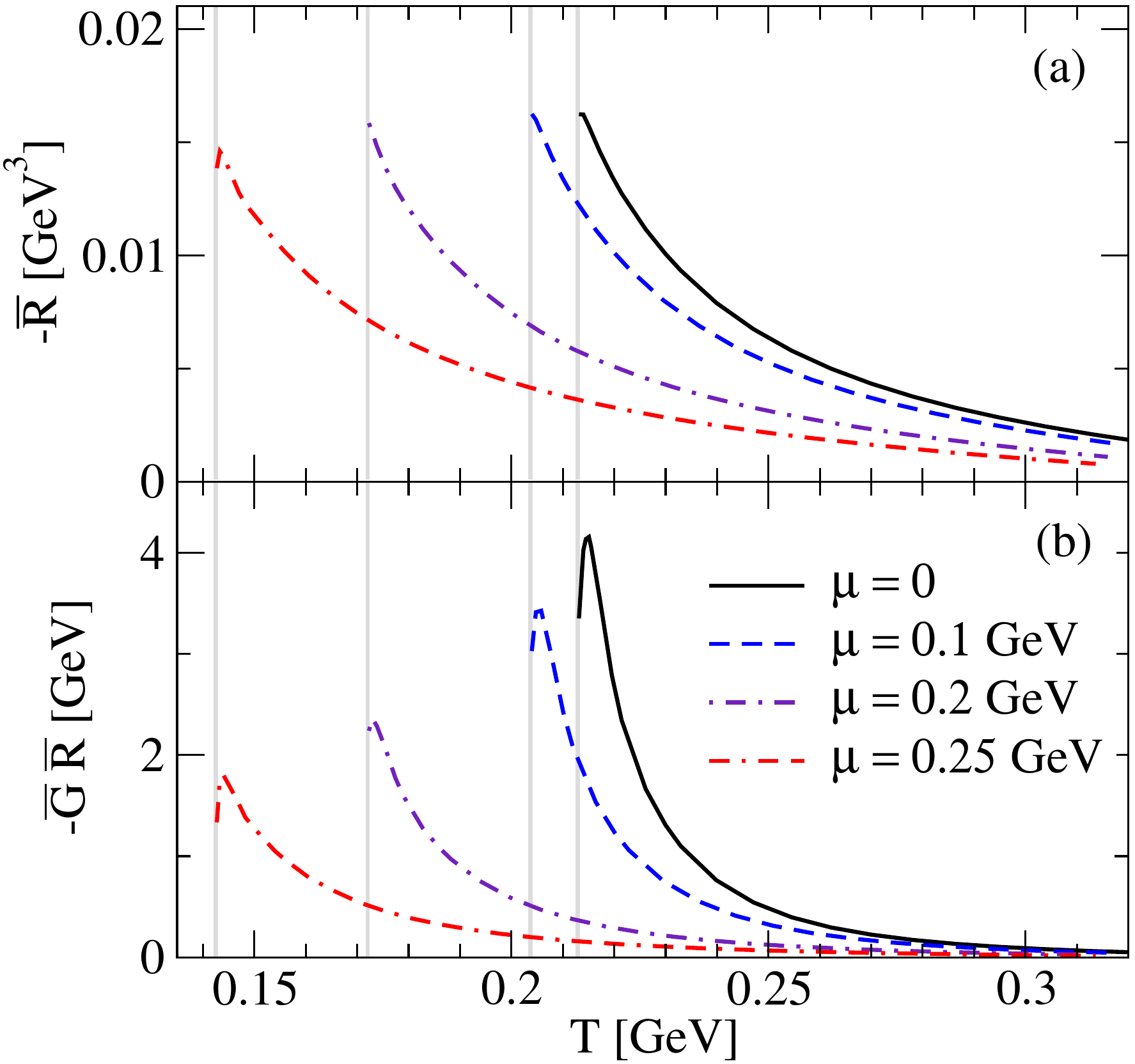}
\caption{ Integral $\bar R$ (a) and the product $\bar G\bar R$ (b) as functions of the temperature for various values of the chemical potential.} 
\label{fig:int_X_G0X_temp} 
\end{center}
\end{figure}
\begin{figure}[!] 
\begin{center}
\includegraphics[width=8.cm,keepaspectratio]{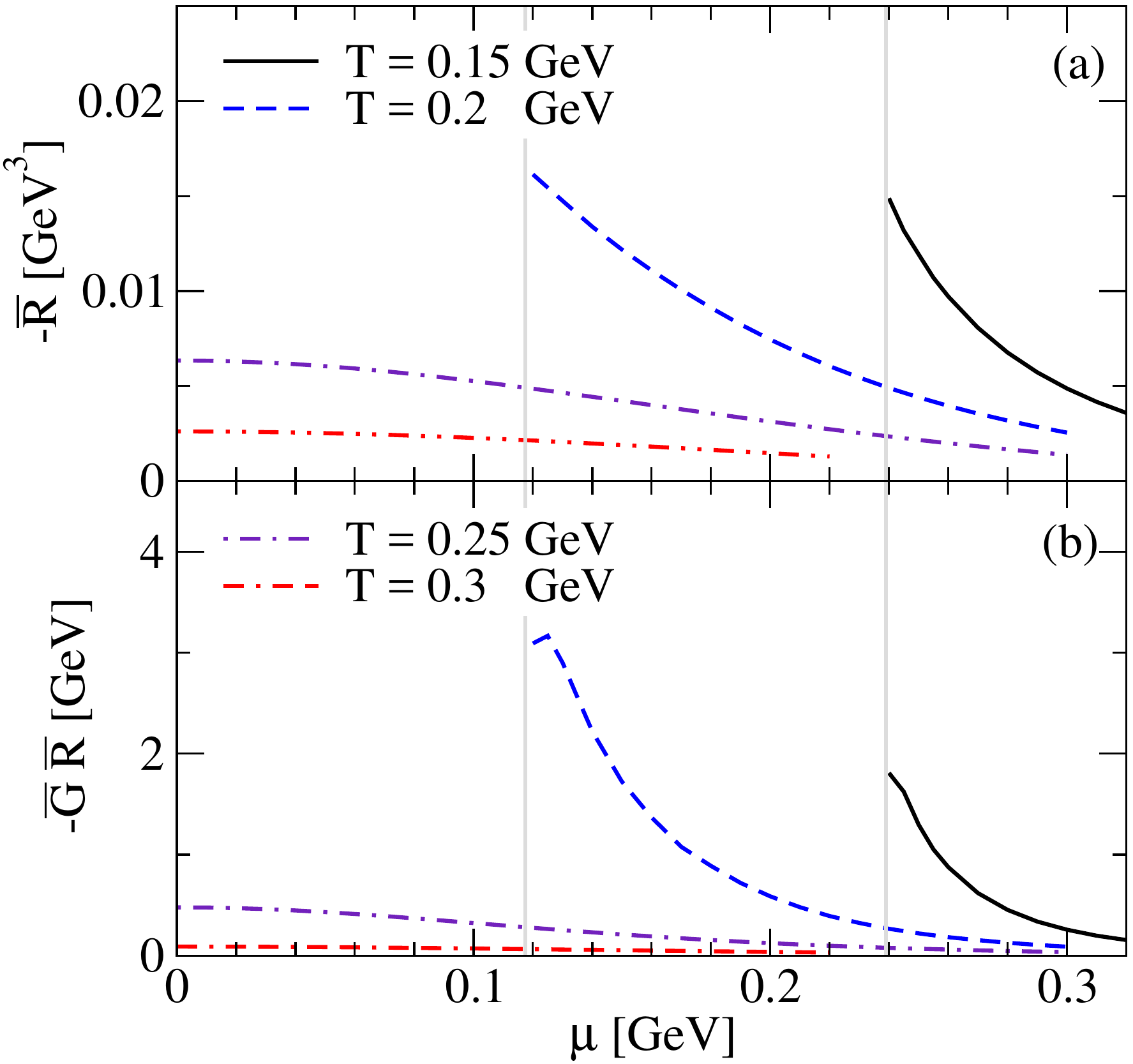}
\caption{ Integral $\bar R$ (a) and the product $\bar G\bar R$ (b) as functions of the chemical potential for various 
values of the temperature. } 
\label{fig:int_X_G0X_mu} 
\end{center}
\end{figure}

 The outer integrands of $R_0$ and
$\bar R$ are rapidly increasing functions of $|\varepsilon|$ for
$|\varepsilon|\leq\Lambda$ and they sharply drop at higher values of
$|\varepsilon|$, as it was the case for the two-dimensional integrals
$I_1$ and $I_2$.
Our analysis
shows that the momentum integrals in Eqs.~\eqref{eq:R_0} and \eqref{eq:R_bar} are invariant under the simultaneous transformations
$\varepsilon\to -\varepsilon$, $\varepsilon'\to -\varepsilon'$,
$\mu\to-\mu$, as expected. Due to this property all integrals are even
functions of the chemical potential.

Figures \ref{fig:int_R0_G0_temp} and \ref{fig:int_R0_G0_mu} illustrate
the temperature and chemical potential dependence of the integral
$R_0$ and the renormalized coupling~\eqref{eq:G_bar0}. The same
dependence for the integral $\bar R$ and the product $\bar G\bar R$ is
shown in Figs.~\ref{fig:int_X_G0X_temp} and \ref{fig:int_X_G0X_mu}.
The latter combination enters the formulas of $\zeta_{1,2}$ components
of the bulk viscosity, see Eq.~\eqref{eq:zeta_12_final}.  It is
remarkable that $R_0$ and $\bar{R}$ remain finite at the Mott
temperature in contrast to the integrals $I_1$ and $I_2$.  The reason
for this behavior can be understood if we recall that at the Mott
temperature the imaginary parts of the self-energies essentially
vanish, therefore the spectral functions transform into
$\delta$-functions:
$A_j(p,\varepsilon)\propto\delta(p^2+m^2-\varepsilon^2)$, where $j$
index labels the Lorentz component. Therefore, the integrands of the
expressions \eqref{eq:R_0} and \eqref{eq:R_bar} will contain a product
of two $\delta$-functions at different arguments.  When integrated
over the variables $\varepsilon$ and $\varepsilon' $, the integral will
consequently have a finite value. (This was not the case for two-dimensional
intergals, where a single energy-integration led to two
$\delta$-functions at the same argument and, therefore, to a divergent
integral.) Apart the different asymptotics for $T\to T_M$, the generic
temperature-density dependence of the three-dimensional intergrals $R_0$ and $\bar R$
does not differ significantly from that of two-dimensional integrals discussed
above. Close to the Mott line we find $R_0\simeq 0.1$ GeV$^2$ and,
therefore, $\bar G\gg G\simeq 10$ GeV$^{-2}$. At high temperatures and
chemical potentials $R_0$ decreases, and $\bar G$ tends to its
``bare'' value. Thus, we conclude that the renormalization of the
coupling constant by multiloop contributions and its effect on the
bulk viscosity should be important in the low-temperature regime close
to the Mott transition line.  We also note that $\bar R$ is alway
negative, which in combination with $I_1<0$ and $I_2>0$ guarantees the
positivity of both components $\zeta_1$ and $\zeta_2$ in the entire
temperature-density range.

\subsection{Bulk viscosities}

\begin{figure}[t] 
\begin{center}
\includegraphics[width=8.cm,keepaspectratio]{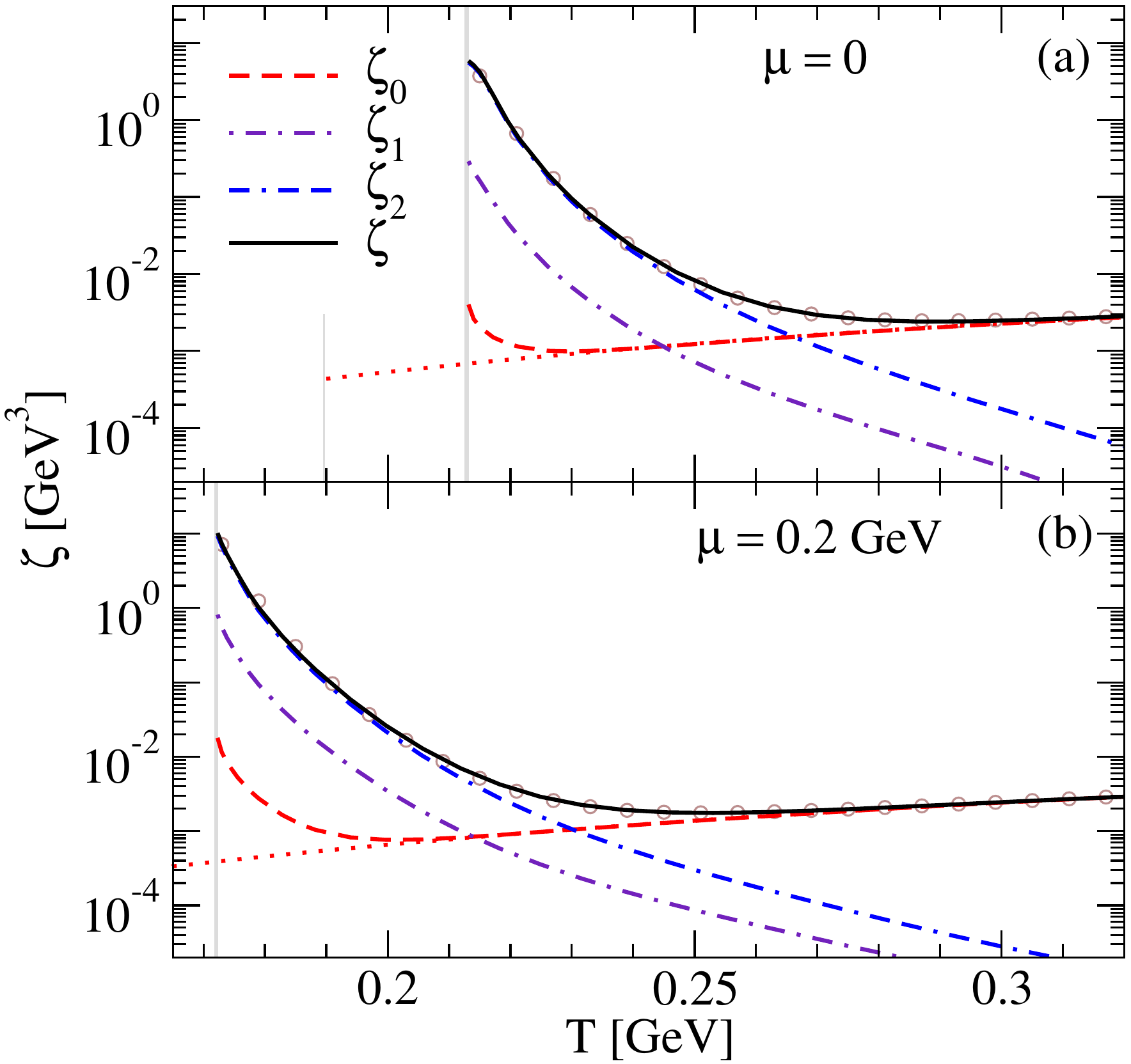}
\caption{ The three components of the bulk viscosity and their sum as functions of the temperature for two values of the chemical potential. The dotted lines correspond to the chiral limit $m_0=0$. The results of the fit formula \eqref{eq:fit_zeta} are shown by circles.} 
\label{fig:zeta_temp} 
\end{center}
\end{figure}

\begin{figure}[!] 
\begin{center}
\includegraphics[width=8.15cm,keepaspectratio]{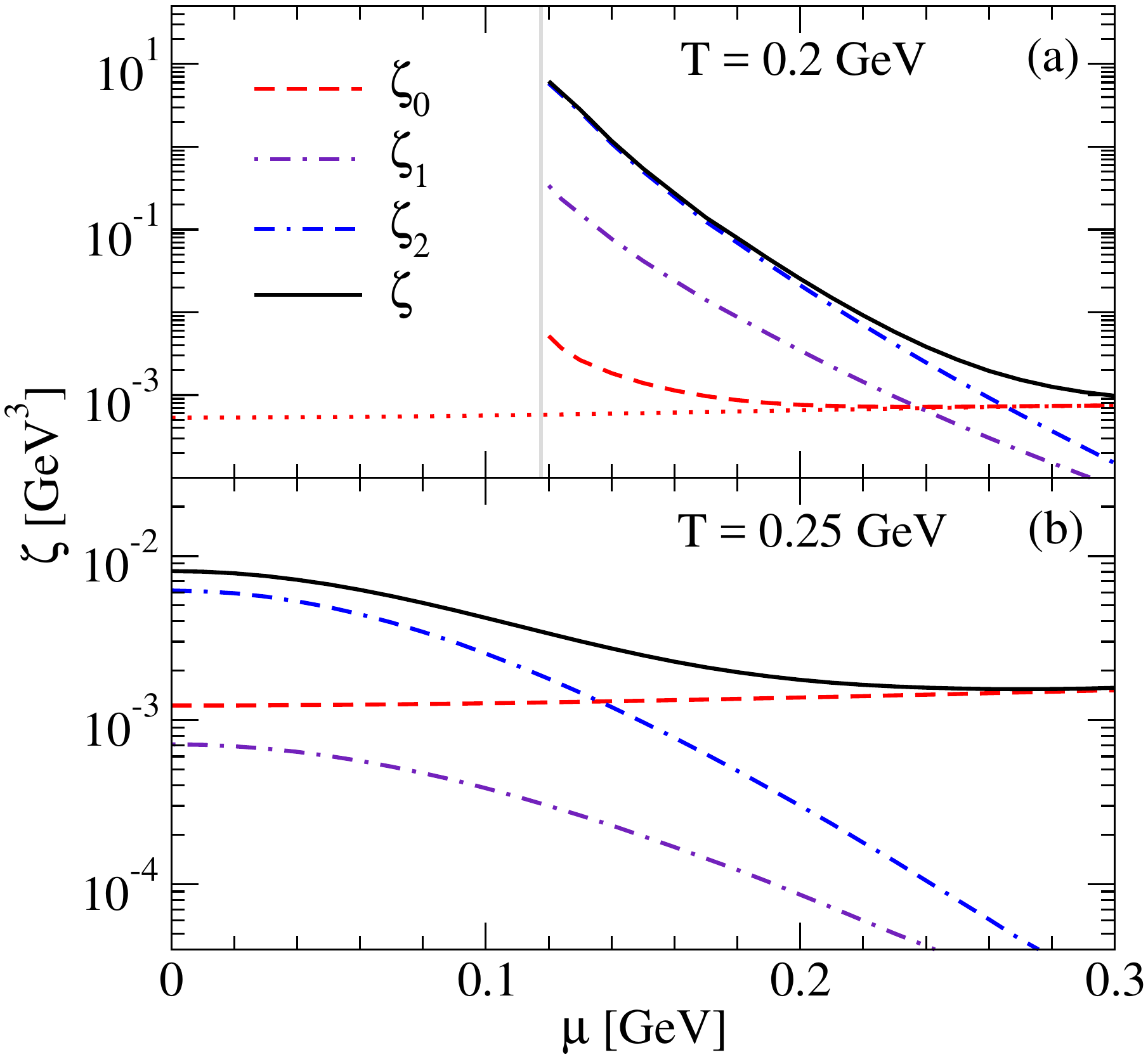}
\caption{ The three components of the bulk viscosity and their sum as functions of the chemical potential for two values of the temperature. The dotted lines correspond to the chiral limit $m_0=0$.} 
\label{fig:zeta_mu} 
\end{center}
\end{figure}

With the analysis above we are in a position to study the behavior of
the components of the bulk viscosity $\zeta_{0}$, $\zeta_{1}$,
$\zeta_{2}$ and their sum $\zeta$. Figures \ref{fig:zeta_temp} and
\ref{fig:zeta_mu} show these quantities as functions of temperature
and chemical potential, respectively.  Because, as we have seen, the
functions $|I_1|$, $I_2$, as well as $R_0$, $\bar G$ and $|\bar R|$
display a maximum at (or close to) the Mott line and decay with
increasing temperature or chemical potential, the multiloop
contributions to the bulk viscosity $\zeta_{1}$ and $\zeta_{2}$ are
expected to show analogous behavior. The one-loop result $\zeta_0$ is
maximal at the Mott line as well, decreases with increasing $T$ or
$\mu$, passes a minimum and increases according to a power law.  At
high temperatures the temperature scaling is $\zeta_0\propto T^3$. This
functional behavior arises from the fact that $\zeta_0$ depends
essentially on the difference of the temporal and vector components of
the spectral function, see Eqs.~\eqref{eq:zeta0_final}--\eqref{eq:abc},
and its asymptotic increase for large $\mu$ or $T$ has been verified
to be the result of the increase of difference between those
components with increasing $T$ or $\mu$, see Fig.~\ref{fig:spectral}.
This is also the reason why the bulk viscosity evaluated in the
one-loop approximation is negligible compared to the shear viscosity,
since the latter depends on the average amplitude of the spectral
functions~\cite{LKW15,2017arXiv170204291H}.

The contribution from the multiloop processes dominates the one-loop
result close to the corresponding Mott line, \ie, at sufficiently low
temperatures or chemical potentials, see Figs.~\ref{fig:zeta_temp} and
\ref{fig:zeta_mu}.  In this regime all three components $\zeta_0$,
$\zeta_1$, $\zeta_2$ and, therefore, also the net bulk viscosity
$\zeta$ drop rapidly with increasing temperature or chemical
potential. The functional behavior of three components of the bulk
viscosity around the Mott line is described by the universal 
formula 
\bea\label{eq:zeta_i} \zeta_i\sim
\exp\left(\frac{a_i}{T/T_M-b_i}\right),\quad i=0,1,2, \eea 
where $a_i$ and $b_i\lesssim 1$ depend only on the chemical
potential. In this regime the following inequalities hold
$\zeta\simeq\zeta_2\gg\zeta_1\gg\zeta_0$, and we see from
Figs.~\ref{fig:zeta_temp} and \ref{fig:zeta_mu} that the one-loop
result $\zeta_0$ underestimates the net bulk viscosity by three orders
of magnitude.

The situation reverses for high $T$ and $\mu$, where the multiloop
contributions $\zeta_1$ and $\zeta_2$ decrease rapidly and one finds
$\zeta\simeq\zeta_0\gg\zeta_2\gg\zeta_1$. As a consequence, the net
bulk viscosity has a mild minimum as a function of temperature. In the
high-temperature regime it increases as $\zeta\propto T^3$, but is
almost independent on the chemical potential.

Thus, we conclude that in the high-$T$ or high-$\mu$ limits the
single-loop approximation correctly represents the bulk viscosity,
\ie, the single-loop provides indeed the leading-order
contribution. This is clearly not the case in the low-$T$ or low-$\mu$
limits, close to the Mott line, where $\zeta_0$ fails to describe
correctly the bulk viscosity, which is dominated by the multiloop
contributions from $\zeta_{2}$.


\subsection{Chiral limit}
\label{sec:chiral}

It is interesting to explore the case when the chiral symmetry is
intact ($m_0=0$). In this case quarks become massless above the
critical (Mott) temperature $T_c$, which implies vanishing multiloop
contributions, as already mentioned in
Sec.~\ref{sec:Kubo_bulk}. Consequently, the bulk viscosity is
determined by the single-loop result \eqref{eq:zeta0_final} with
$m=m_0=0$. As seen from Figs.~\ref{fig:zeta_temp} and
\ref{fig:zeta_mu}, $\zeta_0$ behaves quite
differently in the chiral limit from the case of $m_0\neq 0$ close to the Mott
temperature. It is smooth at the critical temperature and increases
with the temperature by a cubic law in the entire parameter range.  This
behavior can be understood as follows.  At $T\to T_c$ we have $m=0$,
$m_M\to 0$, therefore from
Eqs.~\eqref{eq:im_self}--\eqref{eq:E_min_max_chiral} we find
$\varrho_0 \simeq \varrho_v\to 0$ for high momenta which contribute
mostly to $\zeta_0$. Therefore, from
Eqs.~\eqref{eq:spectral_coeff}--\eqref{eq:N2} we estimate
$n_1\simeq p_0^2-p^2$, $n_2\simeq (p_0^2-p^2)\varrho_0\to 0$ and
$A_{0,v}(p_0, p)\simeq -n_2/(n_1^2+4n_2^2) \sim\delta(p_0^2-p^2)$.
Now substituting $\gamma=1/3$, $\delta=0$ in
Eqs.~\eqref{eq:zeta0_final}--\eqref{eq:abc} (see Appendix~\ref{app:B})
we find that the integrand of $\zeta_0$ is proportional to
$2(\varepsilon^2 A_0-p^2A_v)^2-(\varepsilon^2-p^2)(\varepsilon^2
A_0^2-p^2A_v^2)\sim (\varepsilon^2 -p^2)^2\delta(\varepsilon^2
-p^2)^2\to 0$,
which implies that the integral remains regular in the limit
$T\to T_c$.  In the high-$T$ regime the results for $\zeta_0$ coincide
with those of the case of explicit chiral symmetry breaking.

We note that according to the discussion above $\zeta_0$ component
will vanish in any theory with weakly-interacting massless particles. The
weakness of the interaction implies small spectral widths and,
therefore, nearly on-mass-shell particles with $p=\varepsilon$. As a
consequence, the integrand in Eq.~\eqref{eq:zeta0_final} vanishes, as
expected.

\subsection{Comparison to shear viscosity}

\begin{figure}[t] 
\begin{center}
\includegraphics[width=8.cm,keepaspectratio]{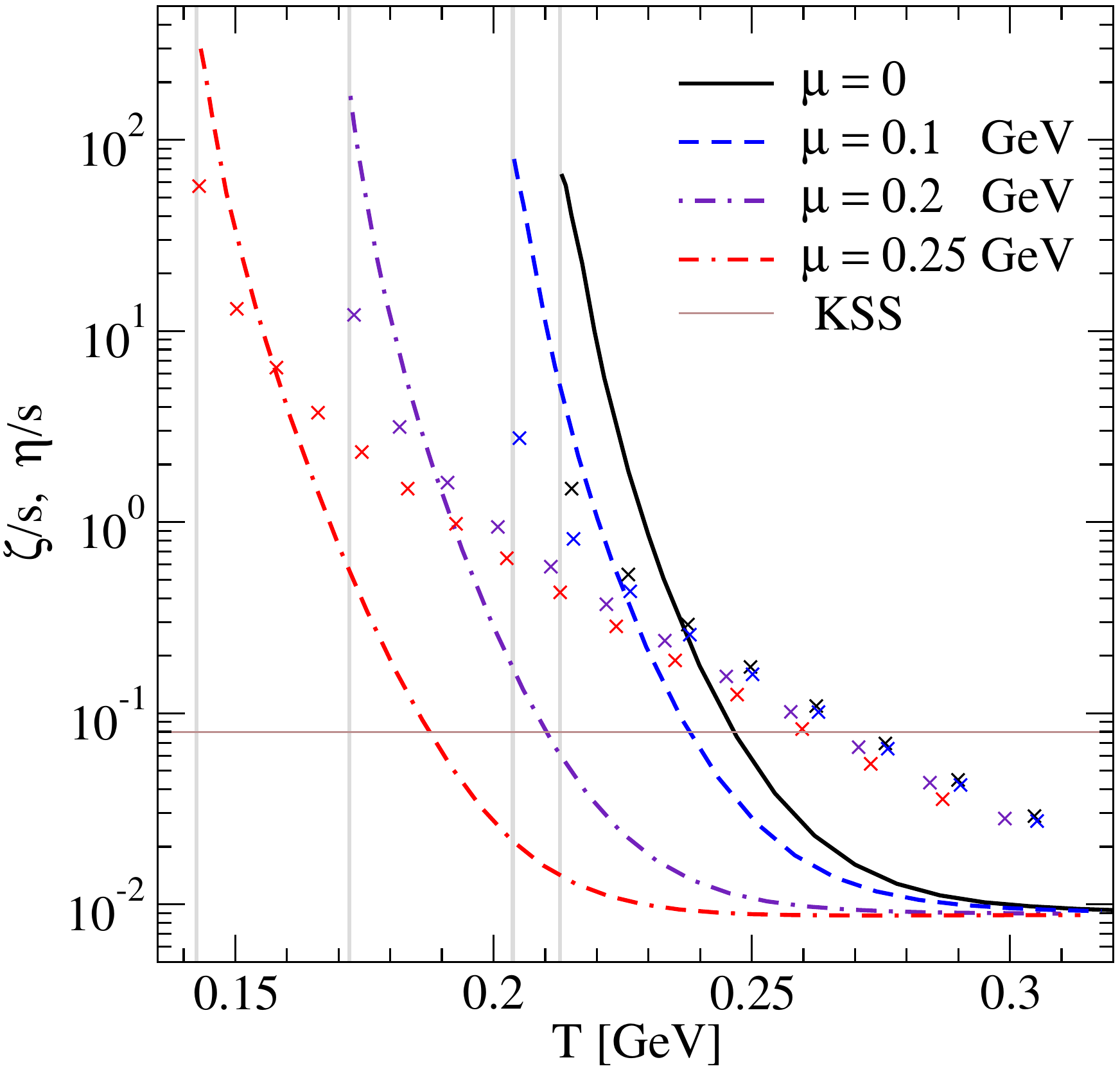}
\caption{ The ratio $\zeta/s$ as function of the temperature for
  several values of the chemical potential. The corresponding $\eta/s$
  ratios are shown for comparison by crosses. The solid horizontal
  line shows the KSS bound~\cite{2005PhRvL..94k1601K}.}
\label{fig:zeta_s_temp} 
\end{center}
\end{figure}
\begin{figure}[!] 
\begin{center}
\includegraphics[width=8.cm,keepaspectratio]{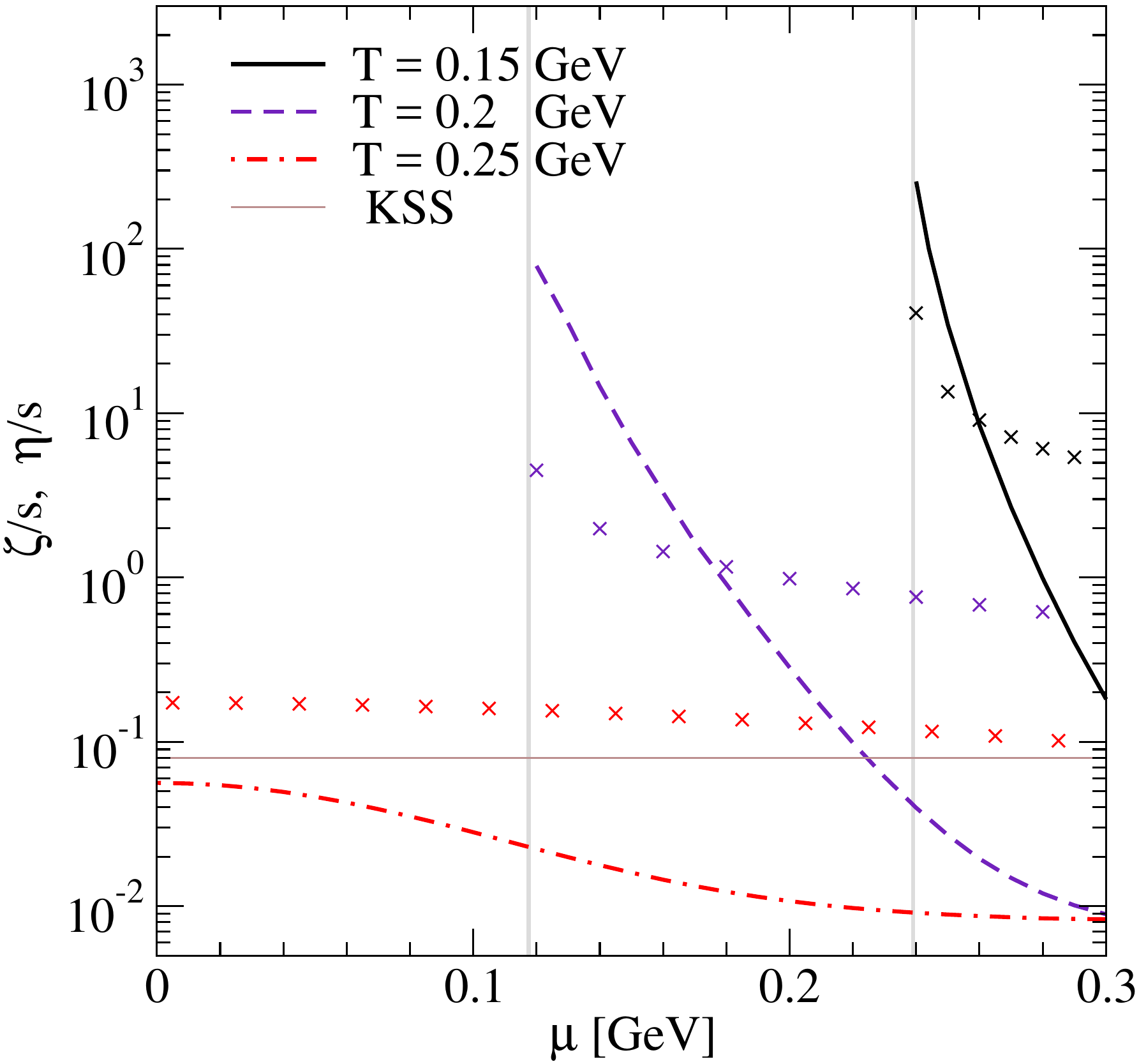}
\caption{ The ratio $\zeta/s$ as function of the chemical potential
  for several values of the temperature. The corresponding ratios
  $\eta/s$ are shown by crosses. The solid horizontal line shows the
  KSS bound~\cite{2005PhRvL..94k1601K}.}
\label{fig:zeta_s_mu} 
\end{center}
\end{figure}

In Figs.~\ref{fig:zeta_s_temp} and \ref{fig:zeta_s_mu} we show the
dependence of the ratio $\zeta/s$ on temperature and chemical
potential, where $s$ is the entropy density, see Appendix \ref{app:B}.
For comparison we show also the ratio $\eta/s$ as computed in
Refs.~\cite{LKW15,2017arXiv170204291H} and the AdS/CFT lower bound $1/4\pi$
on that ratio~\cite{2005PhRvL..94k1601K}. As a general trend, the
ratio $\zeta/s$ increases rapidly close to the Mott transition line
with decreasing temperature or chemical potential and attains its
maximum on this line. It becomes weakly dependent on these quantities
as one moves away from this regime to high-$\mu$ and high-$T$
limit. The $\eta/s$ displays similar behavior, but the increase in the
vicinity of the Mott line is not as steep as for
$\zeta/s$. Numerically we find in this regime $\zeta \ge \eta$ with
$\zeta/\eta \simeq 5 \div 20$ on the Mott line. Thus, we conclude that
close to the Mott transition line {\it the bulk viscosity dominates
  the shear viscosity by large factors and this dominance arises from
  the multiloop processes}. We stress that had we kept only the
one-loop contribution to the bulk viscosity, it would have been
negligible compared to the shear viscosity.  As the temperature or the
chemical potential increases away from the Mott line, $\zeta$
decreases faster than $\eta$ and eventually one reaches the point
where $\zeta = \eta$, beyond which shear viscosity dominates.  This
crossover point appears earlier than the point where
$\zeta_0 \simeq \zeta_2$ beyond which $\zeta_0$ dominates the bulk
viscosity, see the next subsection.  Consequently, we conclude that if
only $\zeta_0$ contribution is kept, then shear viscosity is the
dominant source of dissipation in the entire temperature-density
regime.

In closing we note that in the $T$-$\mu$ region where $\eta$ drops
below the AdS/CFT value $1/4\pi$, the quark-meson fluctuations
considered in this work may not be the dominant processes controlling
the viscous dissipation. Pure gauge fluctuations as well as
quark-quark scattering processes may contribute substantially in
this range of parameters thereby raising the value of $\eta/s$ above
the conjectured bound. 

\subsection{Fits to the bulk viscosity}

The observed nearly universal low-$T$ behavior \eqref{eq:zeta_i} of $\zeta_2$ component with the scaled temperature $T/T_{\rm M}$ for fixed values of the chemical potential
and the high-$T$ asymptotics of $\zeta_0$ suggest fitting the net bulk
viscosity in the whole temperature range by the formula
\bea\label{eq:fit_zeta}
\zeta_{\rm fit}(T,\mu)=a(y)\exp\left[\frac{c(y)}{T/T_M(y)-b(y)}\right]+d(y)T^3,
\eea
with $y=\mu/\mu_0$, where $\mu_0=0.345$ GeV corresponds to the point
where $T_{\rm M}=0$ and the chemical potential attains its maximum on
the Mott line.  The coefficients $a,b,c,d$ are given by
\bea\label{eq:coeff_abc}
a(y) &=& (2.57 - 5.65 y^2)\times 10^{-6}\hspace{0.1cm} [{\rm GeV}^3],\\
b(y) &=& 0.806 - 0.055 y^2 - 0.617 y^4,\\
c(y) &=& 2.89 + 0.96 y^2 +12.73 y^4,\\
d(y) &=& 0.082 + 0.02 y^2.
\eea
The fit formula \eqref{eq:fit_zeta} is valid for chemical potentials
$\mu\le 0.2$ GeV, where its relative error is $\le 10\%$. A comparison
of the fit with the numerical result is given in
Fig.~\ref{fig:zeta_temp}.

In the chiral limit the first term in Eq.~\eqref{eq:fit_zeta}
vanishes, and we are left with pure power-low increase in the whole
temperature-density range
\bea\label{eq:fit_zeta_chiral}
\zeta_{\rm ch}(T,\mu)=T^3(0.082 + 0.168 \mu^2),
\eea
where $T$ and $\mu$ are in GeV units.

We fit also the Mott temperature
displayed in Fig.~\ref{fig:mott_temp1} with the formula 
\bea\label{eq:fit_mott}
\hspace{-0.2cm}
T_{\rm M}^{\rm fit}(\mu)=T_0
\left\{\begin{array}{ll} 1-\sqrt{\gamma y}
e^{-\pi/(\gamma y)}         &0\leq y\leq 0.5,\\
         \sqrt{1.55(1-y)+0.04(1-y)^2}  &0.5< y\leq
                                          1,\end{array}\right.\hspace{-1cm}\nonumber\\
\eea
with $T_0=T_{\rm M}(\mu=0)=0.213$ GeV, and $\gamma =2.7$.  The formula
\eqref{eq:fit_mott} has relative accuracy $\le 3\%$ for chemical
potentials $\mu\le 0.32$ GeV.

Now we define several characteristic temperatures: $T_{\rm min}^0$ and $T_{\rm min}$, corresponding to the minimums of $\zeta_0$ and $\zeta$, respectively;
$T_{02}$ - the temperature of intersection of $\zeta_0$ and $\zeta_2$
components; and $T_{\zeta=\eta}$ - the temperature of intersection of
$\zeta$ and $\eta$. These temperatures vary with the chemical
potential, or, equivalently, with the corresponding value of the Mott
temperature. Interestingly, all these characteristic temperatures turn
out to be linear functions of the Mott temperature with $1\%$ accuracy
and can be fitted as 
\bea\label{eq:fit_temp}
T^*(\mu)=\alpha T_{\rm M}(\mu)+\Delta,
\eea
where $T^*=\{T_{\rm min}^0, T_{\rm min}, T_{02}, T_{\zeta=\eta}\}$,
and the coefficients $\alpha$ and $\Delta$ do not depend on the chemical
potential.  Their numerical values are listed in Table~\ref{tab:1}.
\begin{table}
\begin{tabular}{cccccc}
\hline
$T^*$       &$\quad \alpha\quad$ & $\quad \Delta$ [GeV]\\
\hline 
$T_{\rm min}^0$ & 0.65 & 0.09 \\
$T_{\rm min}$ & 0.86  & 0.106 \\
$T_{02}$ & 0.84  & 0.087 \\
$T_{\eta=\zeta}$  & 1.13  & \hspace{0.1cm}-6$\cdot 10^{-3}$ \\
\hline
\end{tabular}
\caption{The values of the fit parameters in
  Eq.~\eqref{eq:fit_temp}.
}\label{tab:1}
\end{table}

\section{Conclusions}
\label{sec:conclusions}

On the formal side, this work provides a derivation of the bulk
viscosity for relativistic quantum fields in terms of the Lorentz
components of their spectral function within the Kubo-Zubarev
formalism~\cite{1957JPSJ...12..570K,zubarev1997statistical}. It
complements similar expressions for the shear
viscosity~\cite{LKW15} and the electrical and
thermal conductivities~\cite{2017arXiv170204291H} derived earlier.

Practical computations of the bulk viscosity via two-point correlation
function have been carried out within the two-flavor NJL model. The
relevant diagrams were selected by using the $1/N_c$ expansion, where
$N_c$ is the number of colors. 

One of our key results is the observation that the single-loop
contributions, which are dominant for shear viscosity and
conductivities, are insufficient for the evaluation of the bulk
viscosity. We demonstrated that close to the Mott temperature
multiloop contributions, which require resummations of infinite
geometrical series of loops, dominate the one-loop contribution.  We
concentrated on the regime where the dispersive effects arise from
quark-meson scattering above the Mott temperature for decay of mesons
(pions and sigmas) into quarks.  In this regime the bulk viscosity is
a decreasing function of temperature at fixed chemical potential, but
after passing a minimum it increases again. The decreasing behavior is
dominated by multiloop contribution, whereas the high-$T$ increasing
segment is dominated by the one-loop contribution.

Another key result of this study is the observation that the bulk
viscosity dominates the shear viscosity of quark matter in the
vicinity of the Mott temperature by factors of $5\div 20$ depending on the
chemical potential. With increasing temperature the bulk viscosity
decreases faster than the shear viscosity and above a certain
temperature we find $\eta \ge \zeta$. The range of validity of our
comparison is limited by the temperature at which the ratio $\eta/s$
undershoots the KSS bound $1/4\pi$ and obviously the dispersive
effects due to mesonic decays into quarks are insufficient to account for the
shear viscosity of quark matter. Nevertheless, the observation of
large bulk viscosity in the parameter domain of this study may have
interesting and important implications for the hydrodynamical
description of heavy-ion collisions at the RHIC and LHC.

Looking ahead, we anticipate that the formalism
  described here can be straightforwardly extended to include heavier
  flavor quarks, most important being the strange quark. The NJL-model
  Lagrangian can be extended to include vector interactions and/or
  Polyakov loop contributions. As the gluonic degrees of freedom are
  integrated out in the NJL-type models from the outset, the pure
  gauge contributions can be accounted only if one starts with an
  effective model that captures the gauge sector of QCD.

\section*{Acknowledgements}

We thank D. H. Rischke for discussions.  A. H. acknowledges support
from the HGS-HIRe graduate program at Frankfurt University. A. S. is
supported by the Deutsche Forschungsgemeinschaft (Grant No. SE
1836/3-2). We acknowledge the support by NewCompStar COST Action
MP1304.

\appendix

\begin{widetext}

\section{Calculation of the bulk viscosity}
\label{app:A}

Substituting Eq.~\eqref{eq:p_star} into Eq.~(\ref{eq:corp1}) and
taking into account the isotropy of the medium
($[T_{11},T_{33}]=[T_{11},T_{22}]$ etc.)  and the symmetry property of
correlation function in its arguments~\cite{2011AnPhy.326.3075H} we
obtain
\bea\label{eq:corp2} \Pi^{R}_\zeta(\omega) &=& -i\int_{0}^{\infty} dt
e^{i\omega t}\int d\bm r\langle
\frac{1}{3}[T_{11},T_{11}]+\frac{2}{3}[T_{11},T_{22}]
-2\gamma[T_{11},T_{00}]\nonumber\\
&&-2\delta[T_{11},N_{0}]+ 2\gamma\delta
[T_{00},N_{0}]+\gamma^2[T_{00},T_{00}]
+\delta^2[N_{0},N_{0}]\rangle_{0}.  \eea
Further progress requires substituting the explicit expressions for
the components of the energy-momentum tensor \eqref{eq:energymom} and
the particle current \eqref{eq:current} in this expression.  We first
switch to the imaginary time formalism by replacement $t\to -i\tau$
and introduce shorthand notation \bea\label{eq:notations}
\Pi^{(kl)}[\hat{a},\hat{b}](\omega_n)=
\bigg(\frac{G}{2}\bigg)^{k+l-2}\int_{0}^{\beta}d\tau\
e^{i\omega_n\tau}\int d\bm r\langle
T_{\tau}\big((\bar\psi\hat{a}\psi)^k\Big|_{(\bm r,\tau)},
(\bar\psi\hat{b}\psi)^l\Big|_0)\rangle_0, \eea 
where $\hat{a}$ and $\hat{b}$ are either differential operators
(contracted with Dirac $\gamma$ matrices) or interaction vertices
$\Gamma^0_{s/ps}$ appearing in
Eqs.~\eqref{eq:lagrangian}--\eqref{eq:current}.  Then the result of
the substitution can be written as a sum of three terms
\bea\label{eq:corpm1} \Pi^{M}_\zeta(\omega_n) =
\Pi^{M,11}_\zeta(\omega_n) +\Pi^{M,12}_\zeta(\omega_n)
+\Pi^{M,22}_\zeta(\omega_n), \eea where \bea
\label{eq:PiM11}
-\Pi^{M,11}_\zeta(\omega_n)&=&\frac{1}{3}\Pi^{(11)}
[i\gamma_1\partial_1,i\gamma_1\partial_1]+
\frac{2}{3}\Pi^{(11)}
[i\gamma_1\partial_1,i\gamma_2\partial_2]
-2\gamma\Pi^{(11)}
[i\gamma_1\partial_1,-\gamma_0\partial_\tau]\nonumber\\
&-&2\delta\Pi^{(11)}
[i\gamma_1\partial_1,\gamma_0]+2\gamma\delta 
\Pi^{(11)}
[-\gamma_0\partial_\tau,\gamma_0]+
\gamma^2\Pi^{(11)}
[-\gamma_0\partial_\tau,-\gamma_0\partial_\tau]\nonumber\\
&+&\delta^2\Pi^{(11)}[\gamma_0,\gamma_0]+
2(1+\gamma)\Pi^{(11)}
[i\gamma_1\partial_1,i\slashed\partial_\tau-m_0]-
2\gamma(1+\gamma)\Pi^{(11)}
[-\gamma_0\partial_\tau,i\slashed\partial_\tau-m_0]\nonumber\\
&-&2\delta(1+\gamma)\Pi^{(11)}
[\gamma_0,i\slashed\partial_\tau-m_0]+
(1+\gamma)^2\Pi^{(11)}
[i\slashed\partial_\tau-m_0,
i\slashed\partial_\tau-m_0],
\eea
and 
\bea\label{eq:PiM12}
-\Pi^{M,12}_\zeta(\omega_n)&=&
2(1+\gamma)\sum\limits_{\Gamma=\{1,i\bm\tau\gamma_5\}}
\Pi^{(12)}[i\gamma_1\partial_1,\Gamma]-
2\gamma(1+\gamma)\sum\limits_{\Gamma=\{1,i\bm\tau\gamma_5\}}
\Pi^{(12)}[-\gamma_0\partial_\tau,\Gamma]
\nonumber\\
&-&2\delta(1+\gamma)\sum\limits_{\Gamma=\{1,i\bm\tau\gamma_5\}}
\Pi^{(12)}[\gamma_0,\Gamma]+
2(1+\gamma)^2\sum\limits_{\Gamma=\{1,i\bm\tau\gamma_5\}}
\Pi^{(12)}[i\slashed\partial_\tau-m_0,\Gamma],\\
\label{eq:PiM22}
-\Pi^{M,22}_\zeta(\omega_n)&=&
(1+\gamma)^2\sum\limits_{\Gamma,\Gamma'=\{1,i\bm\tau\gamma_5\}}
\Pi^{(22)}[\Gamma,\Gamma'].
\eea
\begin{figure}[t] 
\begin{center}
\includegraphics[width=12cm,keepaspectratio]{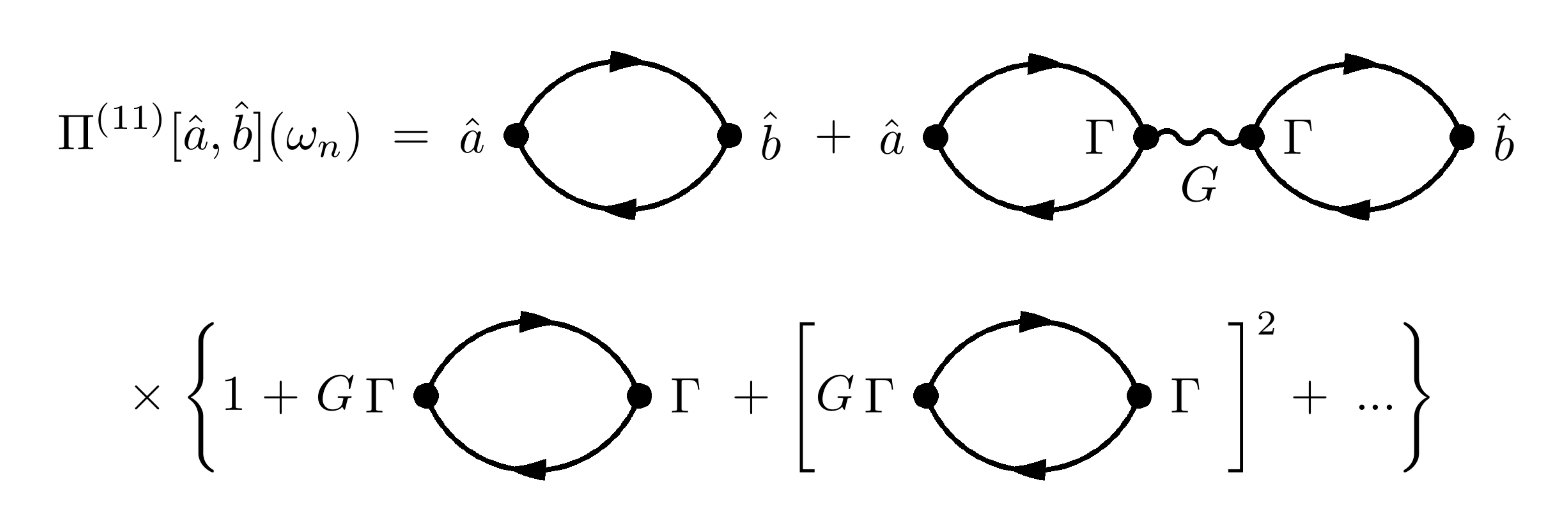}
\caption{ Loop resummation for the correlation function $\Pi^{(11)}[\hat{a},\hat{b}]$ 
defined in Eq.~\eqref{eq:notations} at leading order in $1/N_c$ expansion. }
\label{fig:loops11} 
\end{center}
\end{figure}
\end{widetext} 

\begin{figure}[t] 
\begin{center}
\includegraphics[width=7cm,keepaspectratio]{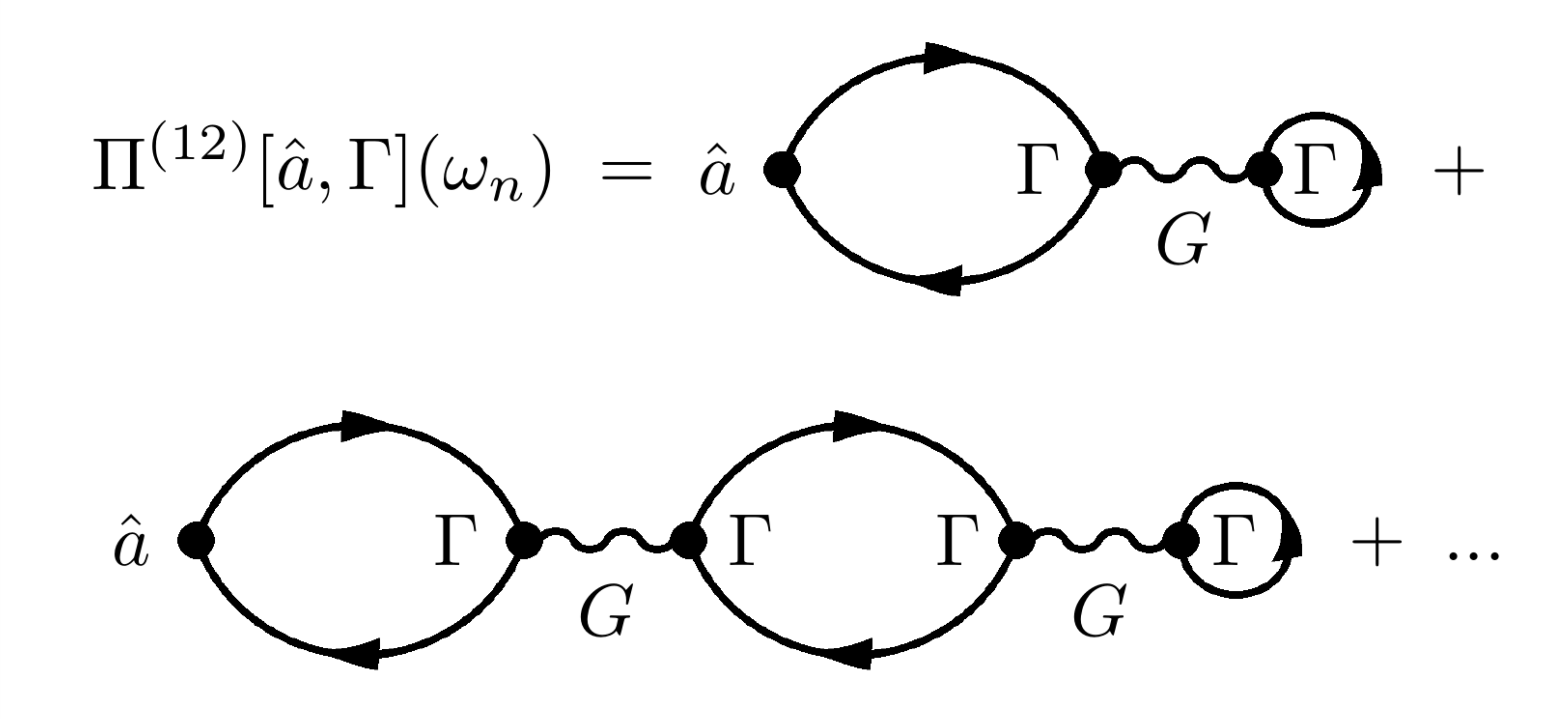}
\caption{ Same as Fig.~\ref{fig:loops11}, but for the function $\Pi^{(12)}[\hat{a},\Gamma$]. }
\label{fig:loops12} 
\end{center}
\end{figure}
\begin{figure}[t] 
\begin{center}
\includegraphics[width=8cm,keepaspectratio]{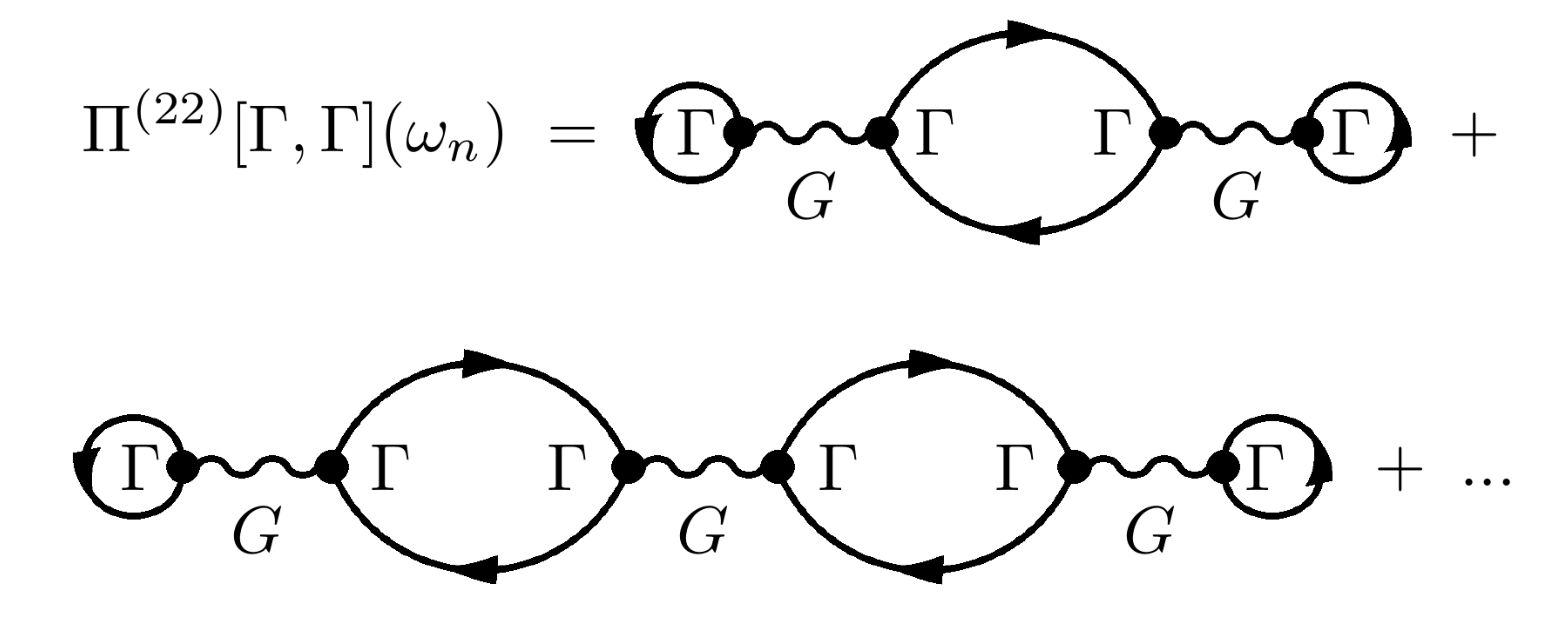}
\caption{ Same as Fig.~\ref{fig:loops11}, but for the function $\Pi^{(22)}[\Gamma,\Gamma$].}
\label{fig:loops22} 
\end{center}
\end{figure}

The three types of correlation functions entering
Eqs.~\eqref{eq:corpm1}--\eqref{eq:PiM22} are shown in Figs.~\ref{fig:loops11}--\ref{fig:loops22}.
Next we note that $\Pi^{(12)}[\hat{a},\Gamma](\omega_n)=
\Pi^{(22)}[\Gamma,\Gamma'](\omega_n)=0,$ because both contain 
 bubble diagrams with one vertex $\Gamma$, which permits only 
 $\omega_n=0$.  Thus, the second and the third terms in Eq.~\eqref{eq:corpm1}
 vanish. We also note that the pseudoscalar vertex with $\gamma_5$
 does not appear in this expression, therefore we are left in all diagrams with the vertex $\Gamma=1$. 

The remaining terms in the two-point correlation
 function can be expressed through the single-loop diagrams given by Eq.~\eqref{eq:ring1_ab} of the main text. With this definition and from Fig.~\ref{fig:loops11} we find that 
\bea\label{eq:cor_11_ab}
\Pi^{(11)}[\hat{a},\hat{b}]=
\Pi_0[\hat{a},\hat{b}]+\tilde{G}
\Pi_0[\hat{a},1]\Pi_0[1,\hat{b}],
\eea
where we introduced a frequency-dependent coupling constant
\bea\label{eq:G_wave}
\tilde{G}(\omega_n)=\frac{G}{1-G\Pi_0[1,1](\omega_n)}.
\eea

To perform the Matsubara sums we need to take into account that the
operators $\hat{a}$ and $\hat{b}$ may depend on $i\omega_l$. (For example,
if $\hat{a}=-\gamma_0\partial_\tau$, in the momentum space we have
$\hat{a}=i\gamma_0\bar{\omega}_l$, $\bar{\omega}_l=\omega_l+\omega_n/2$.) We separate the $i\bar\omega_l$-dependent
parts of these operators by formally factorizing
$\hat{a}...\hat{b}...= f(i\bar\omega_l)\hat{a}_0...\hat{b_0}...$, where
$\hat{a}_0$ and $\hat{b}_0$ are $i\bar\omega_l$-independent parts of
operators $\hat{a}$ and $\hat{b}$. Applying this definition we find 
\bea\label{eq:sums_ab}
S[\hat{a},\hat{b}](\bm p, i\omega_n)&\equiv &
T\sum\limits_l 
\Tr\bigl[\hat{a}G(\bm p, i\omega_l+i\omega_n)
\hat{b}G(\bm p, i\omega_l)\bigr]\nonumber\\
&&\hspace{-2.5cm}=T\sum\limits_l f(i\bar\omega_l)
\Tr\bigl[\hat{a}_0G(\bm p, i\omega_l+i\omega_n)
\hat{b}_0G(\bm p, i\omega_l)\bigr].
\eea
After summation over the Matsubara frequencies and subsequent analytical continuation 
$i\omega_n = \omega_n+i\delta$ we obtain 
\bea\label{eq:resums}
S[\hat{a},\hat{b}](\bm p, \omega)=\int_
{-\infty}^{\infty} d\varepsilon
\int_{-\infty}^{\infty} d\varepsilon'
\Tr[\hat{a}_0A(\bm p, \varepsilon')\hat{b}_0 A(\bm p, \varepsilon)]\nonumber\\
\times 
\frac{\tilde{n}(\varepsilon)f(\varepsilon+\omega/2)-\tilde{n}(\varepsilon')f(\varepsilon'-\omega/2)}
{\varepsilon-\varepsilon' +\omega+i\delta},\hspace{0.5cm}
\eea
where we used the spectral representation 
\bea\label{eq:propagator}
G(\bm p, z)=\int_{-\infty}^{\infty}d\varepsilon
\frac{A(\bm p, \varepsilon)}{z-\varepsilon}.
\eea
This implies that the single-loop polarization tensor is given by 
\bea\label{eq:reloop}
\Pi_0[\hat{a},\hat{b}](\omega)=
\int\frac{d\bm p}{(2\pi)^3}
\int_{-\infty}^{\infty} d\varepsilon 
\int_{-\infty}^{\infty} d\varepsilon'
\Tr[\hat{a}_0 A(\bm p, \varepsilon')
\hat{b}_0 \nonumber\\
\times A(\bm p, \varepsilon)]
\frac{\tilde{n}(\varepsilon')f(\varepsilon'-\omega/2) 
-\tilde{n}(\varepsilon)f(\varepsilon+\omega/2)}
{\varepsilon-\varepsilon' +\omega+i\delta}.
\hspace{0.75cm}
\eea 

The real and imaginary parts of the polarization tensor can now be
computed by applying the Dirac identity. In particular, we find
\bea\label{eq:difresums_ab}
{\rm Im} \Pi_0[\hat{a},\hat{b}](\omega)
\Big|_{\omega=0}=
\frac{d}{d\omega}{\rm Re} \Pi_0[\hat{a},\hat{b}]
(\omega)\Big|_{\omega=0}=0.
\eea
Next we compute from the polarization tensor the relevant structure
needed for the bulk viscosity by defining
\bea\label{eq:difcor_ab}
\frac{d}{d\omega}{\rm Im}\Pi^{(11)}
[\hat{a},\hat{b}](\omega)\bigg\vert_{\omega=0}\hspace{2.5cm}\nonumber\\
=L_0[\hat{a},\hat{b}] + \bar GL_1[\hat{a},\hat{b}]+\bar G^2L_2[\hat{a},\hat{b}],
\eea
where 
\bea
L_0[\hat{a},\hat{b}] &=&
\frac{d}{d\omega}{\rm Im}\Pi_0
[\hat{a},\hat{b}](\omega)\big\vert_{\omega=0},\\
L_1[\hat{a},\hat{b}] &=& 
R_0[\hat{a},1]L_0[1,\hat{b}]
+R_0[1,\hat{b}]L_0[\hat{a},1],\quad\\
L_2[\hat{a},\hat{b}] &=&  L_0[1,1]R_0[\hat{a},1]
R_0[1,\hat{b}],\\
R_0[\hat{a},\hat{b}] &=& {\rm Re}\Pi_0 [\hat{a},\hat{b}]\big\vert_{\omega=0},
\eea
and the effective zero-frequency coupling is given by
\bea\label{eq:G_bar}
\bar{G}\equiv {\rm Re}\bar{G}\big\vert_{\omega=0}
=\frac{G}{1-GR_0[1,1]}.
\eea

Now we calculate the relevant pieces of the polarization for specific
$\hat a$ and $\hat b$ operator combinations. The relevant real 
parts can be written in the generic form
\bea\label{eq:reloop_generic}
R_0[\hat a,\hat b]
&=&-\frac{2N_cN_f}{\pi^4}\!\!
\int_0^\Lambda\!\!\! dp
\int_{-\infty}^{\infty}\!\!\! d\varepsilon
\int_{-\infty}^{\infty}\!\!\! d\varepsilon'\nonumber\\
&&
\frac{\varepsilon^k \tilde{n}(\varepsilon)-\varepsilon'^{k}\tilde{n}(\varepsilon')}
{\varepsilon-\varepsilon'} {\cal O}_R(p,\varepsilon,\varepsilon'),\qquad
\eea
where the factors $N_c=3$ and $N_f=2$ arise from the trace in the color and flavor spaces, respectively; $\Lambda$ is the 3-momentum cutoff parameter. For each specific value of $\hat a$ and $\hat b$ operators we 
have the following functions  ${\cal O}_R$
\begin{subequations}
\bea
R_0\left[1,1\right]  \quad && {\cal O}_R = 
p^2(m^2A_sA'_s+\varepsilon\varepsilon' A_0A_0'\nonumber\\
&&\hspace{2.83cm} -p^2A_vA_v'),\hspace{1cm}
\\
 R_0\left[1,i\gamma_1\partial_1 \right] \quad   
&&{\cal O}_R =  \frac{1}{3}mp^4(A_s'A_v+A_sA_v'),\\
R_0\left[1,\gamma_0\right] \quad &&{\cal O}_R =  mp^2
(\varepsilon A_s'A_0+\varepsilon' A_sA_0'),\\
R_0\left[1,-\gamma_0\partial_\tau\right]
\quad  &&{\cal O}_R =  mp^2
(\varepsilon A_s'A_0+\varepsilon' A_sA_0'),
\eea
and $k=0$ for the first three cases and $k=1$ for the last case.
\end{subequations}

The generic form of the imaginary parts is given by
\bea\label{eq:difimloop_final}
L_0[\hat a,\hat b]= -\frac{2N_cN_f}
{\pi^3}\int_0^\Lambda dp\int_{-\infty}^{\infty} 
d\varepsilon n'(\varepsilon) {\cal O}_I(p,\varepsilon),\quad
\eea
where for each specific value of $\hat a$ and $\hat b$ operators 
the following ${\cal O}_I$ functions should be substituted
\begin{subequations}
\bea\label{eq:difimloop_final}
L_0[1,1]&&  {\cal O}_I = p^2
(m^2A_s^2+\varepsilon^2 A_0^2-p^2 A_v^2),\nonumber\\\\
L_0[1,\gamma_0] &&{\cal O}_I =2m
p^2\varepsilon A_sA_0,\\
L_0[1,-\gamma_0\partial_\tau]
&&{\cal O}_I = 2m
p^2\varepsilon^2 A_sA_0,\\
L_0[1,i\gamma_1\partial_1] &&
{\cal O}_I = \frac{2}{3}mp^4A_sA_v,\\
L_0[\gamma_0,\gamma_0]&&
{\cal O}_I = p^2(m^2A_s^2+\varepsilon^2 A_0^2+p^2A_v^2),
\nonumber\\\\
L_0[-\gamma_0\partial_\tau,-\gamma_0\partial_\tau] &&
{\cal O}_I = p^2\varepsilon^2
(m^2A_s^2+\varepsilon^2 A_0^2+p^2A_v^2),
\nonumber\\\\
L_0[\gamma_0,-\gamma_0\partial_\tau]
&&{\cal O}_I =p^2\varepsilon
(m^2A_s^2+\varepsilon^2 A_0^2+p^2A_v^2),
\nonumber\\\\
L_0[i\gamma_1\partial_1,\gamma_0]
&&{\cal O}_I =\frac{2}{3}p^4\varepsilon A_0A_v,\\
L_0[i\gamma_1\partial_1,-\gamma_0\partial_\tau]&&
{\cal O}_I = \frac{2}{3}p^4\varepsilon^2 A_0A_v,\\
L_0[i\gamma_1\partial_1,i\gamma_1\partial_1]
&&{\cal O}_I =
\frac{p^4}{15}(-5m^2A_s^2+5\varepsilon^2 A_0^2+p^2A_v^2),\nonumber\\\\
L_0[i\gamma_1\partial_1,i\gamma_2\partial_2] 
& &{\cal O}_I =\frac{2}{15}p^6A_v^2.
\eea 
\end{subequations}
With these ingredients the bulk viscosity can be computed by writing
$\zeta=\zeta_0+\zeta_1+\zeta_2$, where the indices on these quantities
match those of the $L$-functions in Eq.~\eqref{eq:difcor_ab}. The
final expressions for these contributions are given by
Eqs.~\eqref{eq:zeta0_final}--\eqref{eq:R_bar} of the main text.

\section{Thermodynamic quantities}
\label{app:B}

In order to find the derivatives in
Eq.~\eqref{eq:gammadelta} we use the relation
$
d\epsilon =Tds+\mu dn,
$
from where we find
\bea\label{eq:gamma_1}
\left(\frac{\partial \epsilon}{\partial p}\right)_n &=&
T\left(\frac{\partial s}{\partial p}\right)_n =
c_V\left(\frac{\partial T}
{\partial p}\right)_n,\\
\label{eq:delta_1}
\left(\frac{\partial n}{\partial p}\right)_\epsilon &=&
-\frac{T}{\mu}\left(\frac{\partial s}
{\partial p}\right)_\epsilon=-\frac{T}{\mu}
\left(\frac{\partial s}{\partial \beta}\right)_\epsilon \left(\frac{\partial \beta}{\partial p}\right)_\epsilon.\quad
\eea
Therefore
\bea\label{eq:gamma_2}
\gamma &=& \left(\frac{\partial p}{\partial \epsilon}\right)_n=\frac{1}{c_V}\left(\frac
{\partial p}{\partial T}\right)_n=-
\frac{\beta^2}{c_V}\left(\frac
{\partial p}{\partial \beta}\right)_n,\\
\label{eq:delta_2}
\delta &=&\left(\frac{\partial p}{\partial n}
\right)_\epsilon = -\beta\mu
\left(\frac{\partial p}{\partial \beta}\right)_\epsilon \left[\left(\frac{\partial s}
{\partial \beta}\right)_\epsilon\right]^{-1},
\eea
where we introduced the heat capacity of a unit volume according to
\bea\label{eq:c_v}
c_V = T\left(\frac{\partial s}{\partial T}\right)_{n}=
-\beta\left(\frac{\partial s}{\partial\beta}\right)_{\mu}
-\beta\left(\frac{\partial s}{\partial\mu}\right)_{\beta}
\left(\frac{\partial\mu}{\partial\beta}\right)_{n}\!\!\!.\quad
\eea
Next we will use the relations
\bea\label{eq:s_deriv}
\left(\frac{\partial s}{\partial \beta}\right)_\epsilon &=&
\left(\frac{\partial s}{\partial\beta}\right)_{\mu}
+\left(\frac{\partial s}{\partial\mu}\right)_{\beta}
\left(\frac{\partial\mu}{\partial\beta}\right)_{\epsilon},\\
\label{eq:p_deriv1}
\left(\frac{\partial p}{\partial \beta}\right)_n &=&
\left(\frac{\partial p}{\partial\beta}\right)_{\mu}
+\left(\frac{\partial p}{\partial\mu}\right)_{\beta}
\left(\frac{\partial\mu}{\partial\beta}\right)_{n},\\
\label{eq:p_deriv2}
\left(\frac{\partial p}{\partial \beta}\right)_\epsilon &=&
\left(\frac{\partial p}{\partial\beta}\right)_{\mu}
+\left(\frac{\partial p}{\partial\mu}\right)_{\beta}
\left(\frac{\partial\mu}{\partial\beta}\right)_{\epsilon}.
\eea
The particle number and entropy densities of quark matter at the leading order in the $1/N_c$ approximation are given by the formulas
\bea\label{eq:quark_number_density}
n&=& \frac{N_cN_f}{\pi^2}\int_0^\infty 
p^2dp[n^+(E_p)-n^-(E_p)],\\
\label{eq:entropy}
s&=&\frac{N_cN_f}{\pi^2}\!\!\int_0^\infty\!\!\! p^2dp
[\beta(E_p-\mu)n^+(E_p) +\beta(E_p+\mu)n^-(E_p) \nonumber\\
&&-\log(1-n^+(E_p))-\log(1-n^-(E_p))],
\eea
with $n^\pm(E)=[e^{\beta(E\mp\mu)}+1]^{-1}$.
The integrals in Eqs.~\eqref{eq:quark_number_density} and \eqref{eq:entropy} are computed according to the
following prescription: no cutoff is imposed as the integral is
convergent, but for momenta $p>\Lambda$ the quark energy is 
taken with the bare mass, \ie,  $E_p=\sqrt{p^2+m_0^2}$.

The internal energy density and the pressure at the leading order are
given by the formulas~\cite{1992RvMP...64..649K,2005PhR...407..205B}
\bea\label{eq:energy_density}
\epsilon&=&\frac{N_cN_f}{\pi^2}
\int_0^{\infty,\Lambda} 
p^2dpE_p[n^+(E_p)+n^-(E_p)-1]\nonumber\\
&&+\frac{(m-m_0)^2}{2G}-C,\\
\label{eq:quark_pressure}
p&=&\frac{N_cN_f}{\pi^2}\int_0^{\infty,\Lambda} 
p^2dp\bigg\{E_p+[n^+(E_p)+n^-(E_p)]\nonumber\\
&&\times \frac{p^2}{3E_p} \bigg\}-\frac{(m-m_0)^2}{2G}+C,
\eea
where the cutoff is applied only for divergent 
parts of the integrals; $C={\rm const}$ and should be fixed by the condition that $p$ and $\epsilon$ vanish in vacuum, \ie, at $T=\mu=0$.

Employing the relations 
\bea\label{eq:fermi_deriv}
\left(\frac{\partial n^\pm}{\partial\beta}\right)_{\mu}&=&
-(E_p\mp \mu) n^\pm(1-n^\pm),\\
\label{eq:fermi_deriv1}
\left(\frac{\partial n^\pm}{\partial\mu}\right)_{\beta}&=&
\pm \beta n^\pm(1-n^\pm),
\eea
and taking the derivatives of Eqs.~\eqref{eq:entropy}, \eqref{eq:quark_pressure}
we obtain  
\bea\label{eq:entropy1}
\left(\frac{\partial s}{\partial\beta}\right)_{\mu}
&=&-\frac{N_cN_f}{\pi^2 T}\int_0^\infty 
p^2dp[(E_p-\mu)^2n^+(1-n^+)\nonumber\\
&& +
(E_p+\mu)^2n^-(1-n^-)],\\
\label{eq:entropy2}
\left(\frac{\partial s}{\partial\mu}\right)_{\beta}
&=&\frac{N_cN_f}{\pi^2 T^2}\int_0^\infty p^2dp
[(E_p-\mu)n^+(1-n^+)\nonumber\\
&& -(E_p+\mu)n^-(1-n^-)],\\
\label{eq:pressure_deriv1}
\left(\frac{\partial p}{\partial \beta}\right)_\mu
&=&-\frac{N_cN_f}{\pi^2}\int_0^\infty 
p^2dp\frac{p^2}{3E_p}
[(E_p-\mu)n^+(1-n^+)\nonumber\\
&& +(E_p+\mu)n^-(1-n^-)],\\
\label{eq:pressure_deriv2}
\left(\frac{\partial p}{\partial \mu}\right)_\beta
&=&\frac{N_cN_f}{\pi^2 T}\int_0^\infty 
p^2dp\frac{p^2}{3E_p}\nonumber\\
&& \times [n^+(1-n^+)-n^-(1-n^-)].
\eea
We neglected the temperature-density dependence
of the constituent quark mass, since this dependence is small above the Mott temperature.
Introducing 
\bea\label{eq:mu_star} \mu^*(\beta,\mu)=
\mu+\beta\left(\frac{\partial\mu}{\partial\beta}\right)_{n},\\
\label{eq:mu_star2}
\mu^{\star}(\beta,\mu)=\mu+\beta\left(\frac{\partial\mu}{\partial\beta}\right)_{\epsilon},
\eea
from Eqs.~\eqref{eq:s_deriv}--\eqref{eq:p_deriv2},
\eqref{eq:entropy1}--\eqref{eq:mu_star2} we obtain
\bea\label{eq:entropy3}
\left(\frac{\partial s}{\partial \beta}\right)_\epsilon 
&=&\mu\frac{N_cN_f}{\pi^2 T}\int_0^\infty p^2dp
[(E_p-\mu^{\star})n^+(1-n^+)\nonumber\\
&&-(E_p+\mu^{\star})n^-(1-n^-)],\\
\label{eq:p_derive_beta_n}
\left(\frac{\partial p}{\partial \beta}\right)_n &=&
 -\frac{N_cN_f}{\pi^2}\int_0^\infty 
p^2dp\frac{p^2}{3E_p}
[(E_p-\mu^*)n^+(1-n^+)\nonumber\\
&&+(E_p+\mu^*)n^-(1-n^-)],\\
\label{eq:p_derive_beta_ep}
\left(\frac{\partial p}{\partial \beta}\right)_\epsilon &=&
-\frac{N_cN_f}{\pi^2}\int_0^\infty 
p^2dp\frac{p^2}{3E_p}
[(E_p-\mu^{\star})n^+(1-n^+)\nonumber\\
&&+(E_p+\mu^{\star})n^-(1-n^-)].
\eea

In order to compute the derivatives
$\left(\partial\mu/\partial\beta\right)_{n}$, $\left(\partial\mu/\partial\beta\right)_{\epsilon}$
we take the $\beta$-derivatives of Eqs.~\eqref{eq:quark_number_density} and \eqref{eq:energy_density} for
$n={\rm const}$ and $\epsilon={\rm const}$, respectively. Since the left-hand-sides vanish trivially, we obtain using Eqs.~\eqref{eq:fermi_deriv}--\eqref{eq:fermi_deriv1}
\bea\label{eq:quark_number_deriv}
\int_0^\infty p^2dp\left[
(E_p-\mu) n^+(1-n^+)-(E_p+\mu) n^-(1-n^-)\right]\nonumber\\
-\beta\left(\frac{\partial\mu}{\partial\beta}\right)_{n}
\int_0^\infty p^2dp\left[ n^+(1-n^+)+ n^-(1-n^-)\right]=0,\nonumber\\
\label{eq:energy_deriv}
\int_0^\infty\!\!\! p^2dp E_p\bigg[
(E_p-\mu) n^+(1-n^+)+(E_p+\mu) n^-(1-n^-)\bigg]\nonumber\\
-\beta \left(\frac{\partial\mu}{\partial\beta}\right)_{\epsilon}
\int_0^\infty p^2dp E_p\left[ n^+(1-n^+)-
 n^-(1-n^-)\right]
=0, \nonumber
\eea
which give in combination with Eqs.~\eqref{eq:mu_star}, \eqref{eq:mu_star2}
\bea\label{eq:quark_number_deriv1}
&&\int_0^\infty \!\!\! p^2dp\bigg[(E_p-\mu^*) 
n^+(1-n^+)\nonumber\\
&&\hspace{2cm}-(E_p+\mu^*) n^-(1-n^-)\bigg]=0,\quad\\
\label{eq:energy_deriv1}
&&\int_0^\infty\!\!\! p^2dp E_p\big[
(E_p-\mu^{\star}) n^+(1-n^+) \nonumber\\
&&\hspace{2cm}+(E_p+\mu^{\star})
n^-(1-n^-)\big]=0.\quad
\eea
The identity \eqref{eq:energy_deriv1} was already used in deriving
Eq.~\eqref{eq:entropy3}.

From Eqs.~\eqref{eq:quark_number_deriv1} and \eqref{eq:energy_deriv1}
we find for $\mu^*$ and $\mu^{\star}$ 
\bea\label{eq:mu_star_1} \mu^*
=\frac{\int_0^\infty p^2dpE_p \left[n^+(1-n^+)-n^-(1-n^-)\right]}{
  \int_0^\infty p^2dp\left[n^+(1-n^+)+
    n^-(1-n^-)\right]},\\
\label{eq:mu_star2_1}
\mu^{\star}=\frac{\int_0^\infty p^2dp E_p^2 
\left[n^+(1-n^+)+n^-(1-n^-)\right]}{
\int_0^\infty p^2dp E_p\left[n^+(1-n^+)-
n^-(1-n^-)\right]}.
\eea

\begin{figure}[t] 
\begin{center}
\includegraphics[width=8.cm,keepaspectratio]{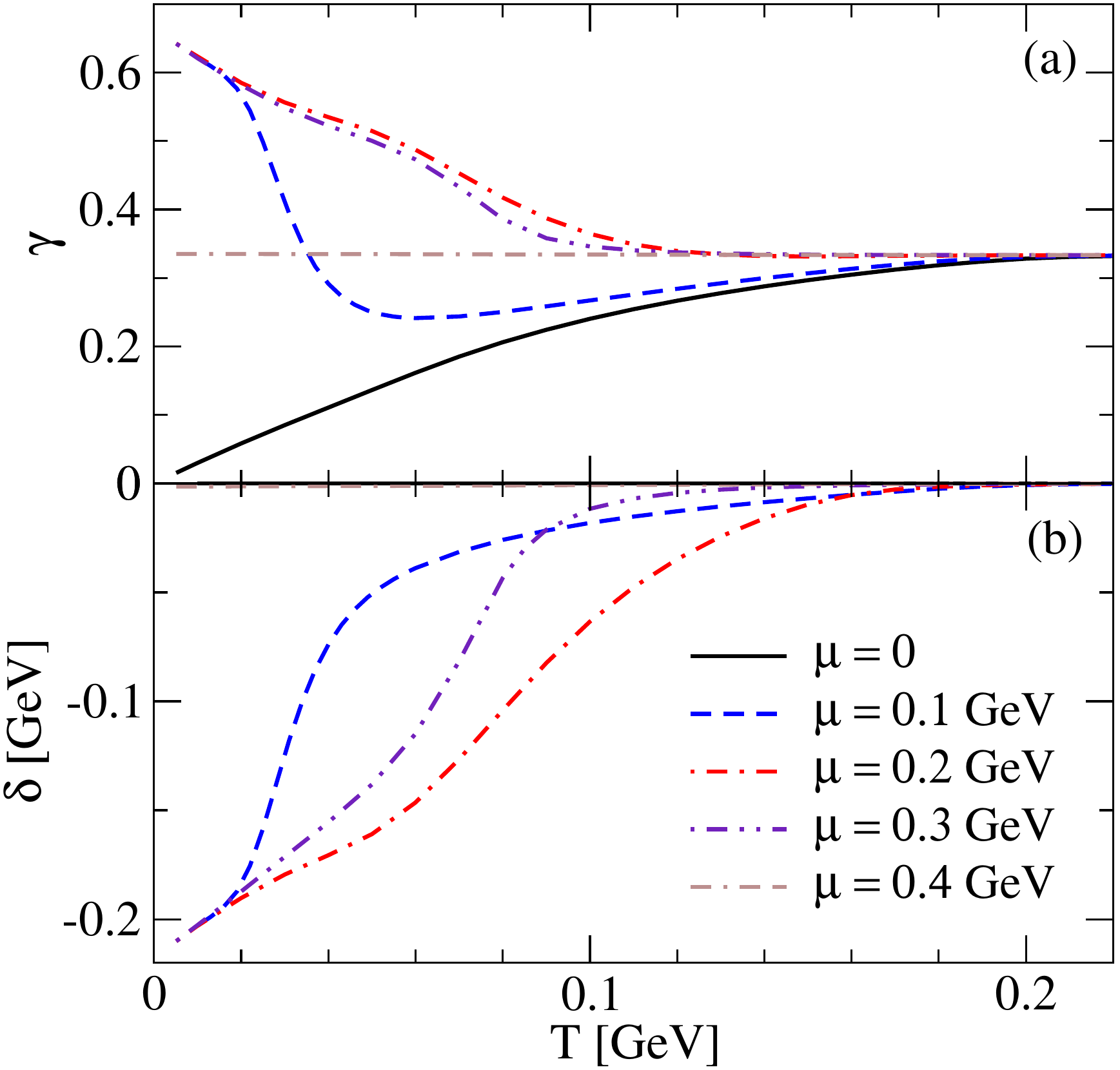}
\caption{ The coefficients $\gamma$ (a) and $\delta$ (b) as functions
  of the temperature for various values of the chemical potential.}
\label{fig:gamma_delta} 
\end{center}
\end{figure}

Using Eqs.~\eqref{eq:c_v}, \eqref{eq:entropy1}, \eqref{eq:entropy2},
\eqref{eq:mu_star} and \eqref{eq:quark_number_deriv1} we find for the
heat capacity
\bea\label{eq:c_v2}
c_V=\frac{N_cN_f}{\pi^2 T^2}\int_0^\infty
p^2dpE_p[(E_p-\mu^*)n^+(1-n^+) \nonumber\\
 +(E_p+\mu^*)n^-(1-n^-)].
\eea

\begin{widetext}
Finally, substituting Eqs.~\eqref{eq:entropy3}--\eqref{eq:p_derive_beta_ep} and \eqref{eq:c_v2}
into Eqs.~\eqref{eq:gamma_2} and \eqref{eq:delta_2} we obtain
\bea\label{eq:gamma}
\gamma &=&\frac{\int_0^\infty p^4dp(3E_p)^{-1}
[(E_p-\mu^*)n^+(1-n^+)+(E_p+\mu^*)n^-(1-n^-)]}
{\int_0^\infty p^2dpE_p[(E_p-\mu^*)n^+(1-n^+)
+(E_p+\mu^*)n^-(1-n^-)]},\\
\label{eq:delta}
\delta &=& \frac{\int_0^\infty p^4dp
(3E_p)^{-1}[(E_p-\mu^{\star})n^+(1-n^+)+
(E_p+\mu^{\star})n^-(1-n^-)]}{\int_0^\infty p^2dp
[(E_p-\mu^{\star})n^+(1-n^+)-
(E_p+\mu^{\star})n^-(1-n^-)]}.
\eea
\end{widetext}

Equations~\eqref{eq:mu_star_1}, \eqref{eq:mu_star2_1}, \eqref{eq:gamma} and
\eqref{eq:delta} imply that $\mu^*, \mu^\star$ and $\delta$ are odd
and $\gamma$ - even functions of the chemical potential.  The
thermodynamic quantities $\gamma$ and $\delta$ given by
Eqs.~\eqref{eq:gamma} and \eqref{eq:delta} are shown in
Fig.~\ref{fig:gamma_delta}. We find that $\gamma$ tends to a constant
value $\gamma= 1/3$ at high temperatures and chemical potentials. In $T\to 0$ limit $\gamma\to 0$ for $\mu=0$ and $\gamma\to 2/3$ for
intermediate values of the chemical potential $T\ll\mu<m(T=0)$.  Note
that in the limit of vanishing chemical potential $\gamma =s/c_V$
coincides with the sound speed, which makes clear the high-temperature
asymptotics of $\gamma$.  We find also that $\delta$ is numerically
negligible compared to the typical energy scales for the whole
temperature-density range of interest. It vanishes asymptotically at
high temperatures and densities, but tends to a constant limit
$\delta\to -2m(T=0)/3\simeq 0.22$ GeV at $T\to 0$ if
$m(T=0)>\mu\gg T$.  In the chiral limit $m=0$ above the critical
temperature $T_c$, and we find from Eqs.~\eqref{eq:energy_deriv1},
\eqref{eq:gamma} and \eqref{eq:delta} constant values $\gamma=1/3$ and
$\delta=0$.

\providecommand{\href}[2]{#2}\begingroup\raggedright\endgroup

\end{document}